\documentclass[12pt]{article}

\usepackage[a4paper]{geometry}
\usepackage{graphicx}
\usepackage{amsmath}
\usepackage{natbib}
\usepackage{algorithm}
\usepackage{url}
\usepackage{hyperref}
\usepackage{authblk}

\title{Statistical Design and Analysis for Robust Machine Learning: A Case Study from COVID-19}
\author[1]{Davide Pigoli\footnote{These authors contributed equally to this work.}\footnote{Address for correspondence: Department of Mathematics, King's College London, Strand, London WC2R 2LS, United Kingdom. Email: davide.pigoli@kcl.ac.uk.}}
\author[1]{Kieran Baker$^*$}
\author[2]{Jobie Budd}
\author[3]{Lorraine Butler}
\author[4]{Harry Coppock}
\author[3]{Sabrina Egglestone}

\author[1]{Steven G.\ Gilmour}
\author[5]{Chris Holmes}
\author[3]{David Hurley}

\author[6]{Radka Jersakova}

\author[7]{Ivan Kiskin}

\author[1]{Vasiliki Koutra}

\author[3]{Jonathon Mellor}

\author[8]{George Nicholson}

\author[3]{Joe Packham}

\author[9]{Selina Patel}

\author[3]{Richard Payne}

\author[5]{Stephen J.\ Roberts}

\author[10]{Bj\"{o}rn W.\ Schuller}

\author[11]{Ana Tendero-Ca$\tilde{\mathrm{n}}$adas}
 \author[12]{Tracey Thornley}

\author[3]{Alexander Titcomb}

\affil[1]{\normalsize King's College London, UK and The Alan Turing Institute, London, UK.}
\affil[2]{\normalsize University College London, UK.}
\affil[3]{\normalsize UK Health Security Agency, London, UK.}
\affil[4]{\normalsize Imperial College London, UK.}
\affil[5]{\normalsize University of Oxford, UK and The Alan Turing Institute, London, UK.}
\affil[6]{\normalsize The Alan Turing Institute, London, UK.}
\affil[7]{\normalsize University of Surrey, UK.}
\affil[8]{\normalsize University of Oxford, UK.}
\affil[9]{\normalsize UK Health Security Agency, London, UK and University College London, UK.}
\affil[10]{\normalsize The Alan Turing Institute, London, UK and Imperial College London,  UK.}
\affil[11]{\normalsize UK Health Security Agency, London, UK and 
University of Brighton, UK.} 
\affil[12]{\normalsize University of Nottingham, UK.}

\date{}

\begin{document}

\maketitle

\newpage

\begin{abstract}

Since early in the coronavirus disease 2019 (COVID-19) pandemic, there has been interest in using artificial intelligence methods to predict COVID-19 infection status based on vocal audio signals, for example cough recordings. However, existing studies have limitations in terms of data collection and of the assessment of the performances of the proposed predictive models. This paper rigorously assesses state-of-the-art machine learning techniques used to predict COVID-19 infection status based on vocal audio signals, using a dataset collected by the UK Health Security Agency. This dataset includes acoustic recordings and extensive study participant meta-data. We provide guidelines on testing the performance of methods to classify COVID-19 infection status based on acoustic features and we discuss how these can be extended more generally to the development and assessment of predictive methods based on public health datasets.

\end{abstract}


\section{Introduction}

From the beginning of the coronavirus disease 2019 (COVID-19) pandemic, it has been recognised that rapid and widespread testing is one of the most important public health measures for containing the spread of the virus \citep{who2020}. The gold-standard reverse-transcription polymerase chain reaction (RT-PCR) test is very sensitive and specific to severe acute respiratory syndrome 2 (SARS-CoV-2) viral ribonucleic acid (RNA), but slow and expensive to carry out, so is not considered practical for widespread community testing. Lateral flow antigen tests offer a faster way to identify COVID-19 positive individuals, especially those with high viral load. However, they have cost and usability challenges and are less sensitive than a PCR test. With the continuing impact of COVID-19, there is a requirement for faster, simpler and cheaper ways to test for infection to reduce the impact of widespread transmission.

One of the earliest identified symptoms of COVID-19 was a distinctive dry cough. Since 2020, researchers \citep[see, e.g.,][]{laguarta2020covid, han2021exploring, brown2020exploring} have explored the use of machine learning to classify forced cough samples into those from COVID-19 positive and COVID-19 negative individuals. They have reported results which indicate high levels of accurate classification, however,
it is difficult to assess from these studies how well the classifiers might actually perform in practice, as discussed in \cite{coppock2021covid}.

In early 2021, a study was set up by the UK Health Security Agency (UKHSA)
and the \href{https://www.turing.ac.uk/research/research-projects/turing-rss-health-data-lab}{Turing-RSS Health Data Laboratory} -- a working partnership between The Alan Turing Institute and Royal Statistical Society -- to rigorously assess the feasibility of these methods as a public health tool. A team of statisticians and computer scientists was brought together, with the aim of assessing what levels of accuracy might be achievable in practice. An analysis of subject-level covariates (meta-data) of the available data was used to choose training and test sets to enable development and assessment of machine learning models with a minimum of bias and as great a chance as possible to have a similar performance when applied to the wider population.

A relevant issue when estimating the accuracy of machine learning methods on public health datasets of observational nature, or surveys with a high level of non-responses, is the presence of bias in the dataset, i.e.,
the dataset does not reflect the population of interest in some important aspect. This can lead to estimates of accuracy that cannot be replicated when the methods are later applied in practice. A second issue is the presence of confounders, i.e., variables that are correlated with the outcome and the predictors and that the machine learning method can learn to predict instead of the actual outcome of interest to obtain a greater accuracy within the dataset. It is doubly problematic when bias and confounding are both present, in the sense that confounding variables are correlated with the outcome of interest in the dataset but not in the general population. This can lead to a machine learning method that is extremely accurate in the dataset but useless for practical purposes.

While the focus on this work is on the prediction of COVID-19 infection status from acoustic features, many of the statistical issues discussed in this paper are valid more generally for machine learning methods trained on observational data, and the approach we suggest for the assessment of these methods in Section \ref{sec:choice} is an important contribution to the currently active research on explainable AI \citep{watson2022conceptual,rudin2019stop}, in particular for healthcare applications \citep[see, for example, ][]{babic2021beware}.

\section{Previous work on vocal acoustic features for predicting COVID-19 status }
\label{sec:review}

This section provides a discussion of some of the previously published papers which have sparked interest in the use of acoustic biomarkers to predict COVID-19 infection status and demonstrated seemingly promising results in this direction.

\citet{laguarta2020covid} used the MIT Open Voice model to predict COVID-19 infection status based on cough recordings, using crowd-sourced data collected from a web form (5320 subjects), where COVID-19 infection status was mostly self-reported. 
The authors report a sensitivity of $100\%$ on asymptomatic patients (of the group of subjects diagnosed with an ``official test''), but it is unclear whether asymptomatic patients are properly held out for out-of-sample prediction on the test set. With a corresponding specificity of $83.2\%$, it is possible that the $100\%$ sensitivity is so high due to over-fitting, rather than high predictive power. The absence of demographic information (e.g., on geographical region or age) also makes it difficult to assess the generalisability of the results.

\citet{brown2020exploring} collected a crowdsourced dataset comprised of submissions from 6613 participants through a web-based and a mobile app (COVID-19 Sounds), and predicted COVID-19 status based both on hand-crafted features extracted from coughs and breathing sounds and features extracted using transfer learning techniques from the same signals, using various machine learning techniques, such as logistic regression, gradient boosting trees and Support Vector Machines (SVMs). They reported an area under the curve of the Receiver Operating Characteristic curve (ROC-AUC) of $80\%$. Their use of transfer learning and handcrafted features therefore appeared to show promise. However, the number of positive cases in the dataset is small, especially when considering test/validation sample sizes. In addition, the authors selected negative cases from countries with low infection prevalence at the time of sampling, which may introduce bias associated with the mother tongue (through its influence on speech physiology) that is not accounted for. To ensure the classifier is not just classifying participants to their country, a more appropriate approach may have been to select a group of COVID-19 negative participants within a high-prevalence country. The effect of demographics on the data and sound samples is not considered, perhaps due to the limitations of the dataset. Also, people in the age bracket 20 to 49 years old seem to be over-represented in the sample. Moreover, the imbalance between positive and negative cases suggests that the use of the ROC curve might not be suitable, and adjustments or alternative methods might be appropriate. 

\citet{han2021exploring} discussed voice-based models for COVID-19 status that used symptoms to classify positive and negative cases. The authors combined symptom covariates with speech recordings. However, the performance greatly varied across folds (wide standard deviation), which indicates epistemic uncertainty and the requirement of more training data. It was noted that a much smaller dataset was used compared to \citet{brown2020exploring} (with 343 participants coming from 4 countries, unequally represented). The authors implemented the synthetic minority oversampling technique (SMOTE), which adds synthetic data to the minority class to achieve balance between the positive and negative cases. There is no explicit information related to the way the positive cases were chosen. Moreover, similar to \citet{brown2020exploring}, there is a lack of evaluation related to demographics. More recently, this issue was addressed by \cite{han2022sounds}, where it is indeed shown how biases and participant splits can affect the performance of the method, using a dataset of 2478 volunteers who self-reported COVID-19 test results.

To summarise,  there are reports of accurate COVID-19 status prediction from vocal acoustic features across the existing literature. However, there is also a recognition of the need for large, clinically referenced datasets with sufficient metadata and transparency when using machine learning and artificial intelligence techniques for diagnostics. As mentioned above, \citet{han2022sounds} discuss how the evaluation of these diagnostic procedures can be affected by biases in the dataset. As we argue in this paper, a careful design of the assessment procedure is also of paramount importance, and is only possible when accurate metadata from study participants are available. We are going to show that, when the study (i.e., both the data collection mechanisms and the procedure to estimate and compare methods' performance) is designed specifically to assess out-of-sample generalisability and to investigate the predictive power of the acoustic information, the results may no longer be as impressive.

\section{Data Collection}


Developing and accurately assessing bioacoustics-based classification methods requires a carefully designed study, both from the point of view of the data collection and from the point of view of how to test and compare predictive performances. Concerning the latter, we do not
need to rely on a randomised controlled trial, since it is possible to compare classification methods using all the subjects and there is no need to rely on methods which require counterfactuals. However, an important question is then how the performance of these methods extends
to the population of interest, as opposed to the collected dataset.

Therefore, the main issue is to ensure that data are collected from a well understood environment, with as much relevant metadata as possible, to ensure that possible confounding variables are accounted for. In principle, we would like the dataset to be representative of the population of interest, but in practice we expect some biases to be present and collecting all the relevant meta-data allows us to adjust the assessment procedure to account for these biases. For example, the split of the data into training and test sets should be done in such a way as to guarantee that the true out-of-sample performance can be measured as well as possible.

Following the publication of initial studies reporting accurate classification of COVID-19 infection status from vocal and respiratory audio, the UK Health Security Agency (UKHSA, formerly NHS Test and Trace, the Joint Biosecurity Centre, and Public Health England) were commissioned to collect a dataset to allow for the independent evaluation of these studies.
The ``Speak up and help beat coronavirus'' study \citep{linktodhscpage} was set up to collect data for this purpose.  A description of the resulting dataset (UK COVID-19 Vocal Audio Dataset) and how to access it can be found in \citet{ciab_data}. Here, 
we give a quick overview and highlight the key variables used in the design and evaluation of the predictive procedure described and assessed in detail in \cite{ciabdraft}. 
The primary analysis that we discuss in this paper was based on data collected from this study between 01 March 2021 and 29 November 2021, giving 39850 submissions. It should be noted that the dataset described in  \citet{ciab_data} also contains  submissions collected after 29 November 2021. However, this was the date where the meta-data were analysed to design the procedure to assess the methods' performances. Thus, data collected after this cut-off date were not considered in the primary analysis, but they were used for some of the follow-up analysis described in \cite{ciabdraft}.

Volunteer participants recruited via two routes: (i) NHS Test and Trace community testing in England and (ii) the REACT-1 (REal-time Assessment of Community Transmission) survey \citep{riley2020real}.  In the period covered by the data collection, people were advised to seek a PCR-test through NHS Test and Trace if they were experiencing COVID-19 symptoms, they were identified as a close contact of a positive case, or following a positive rapid antigen (lateral flow) test (until 11th January 2022). On the other hand, REACT-1 was commissioned by the UK Department of Health and Social Care to estimate the prevalence of SARS-CoV-2 infection in the community in England (and influenza A and B in later survey rounds). It was carried out by Imperial College London in partnership with Ipsos MORI using repeat, random, cross-sectional sampling of the population.

\begin{figure}
  \centering
  \begin{tabular}{ | p{1.9cm} | c | c | c | }
    \hline
    & Missing audio (1213) & Missing meta-data (162) & Final dataset (37018) \\ \hline
    Gender & 
    \begin{minipage}{.27\textwidth}
      \includegraphics[width=\linewidth]{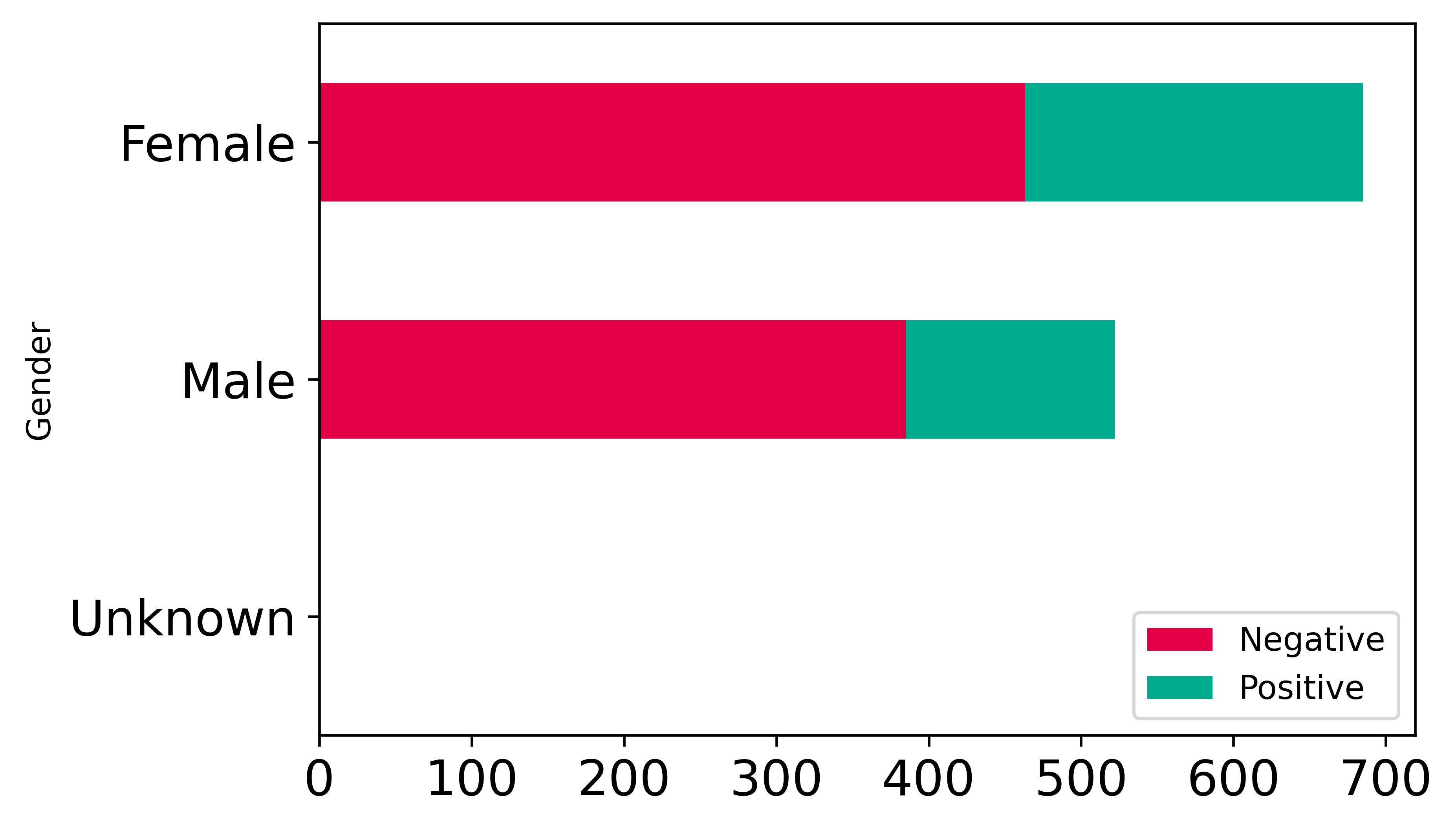} 
    \end{minipage} & 
    \begin{minipage}{.27\textwidth}
      \includegraphics[width=\linewidth]{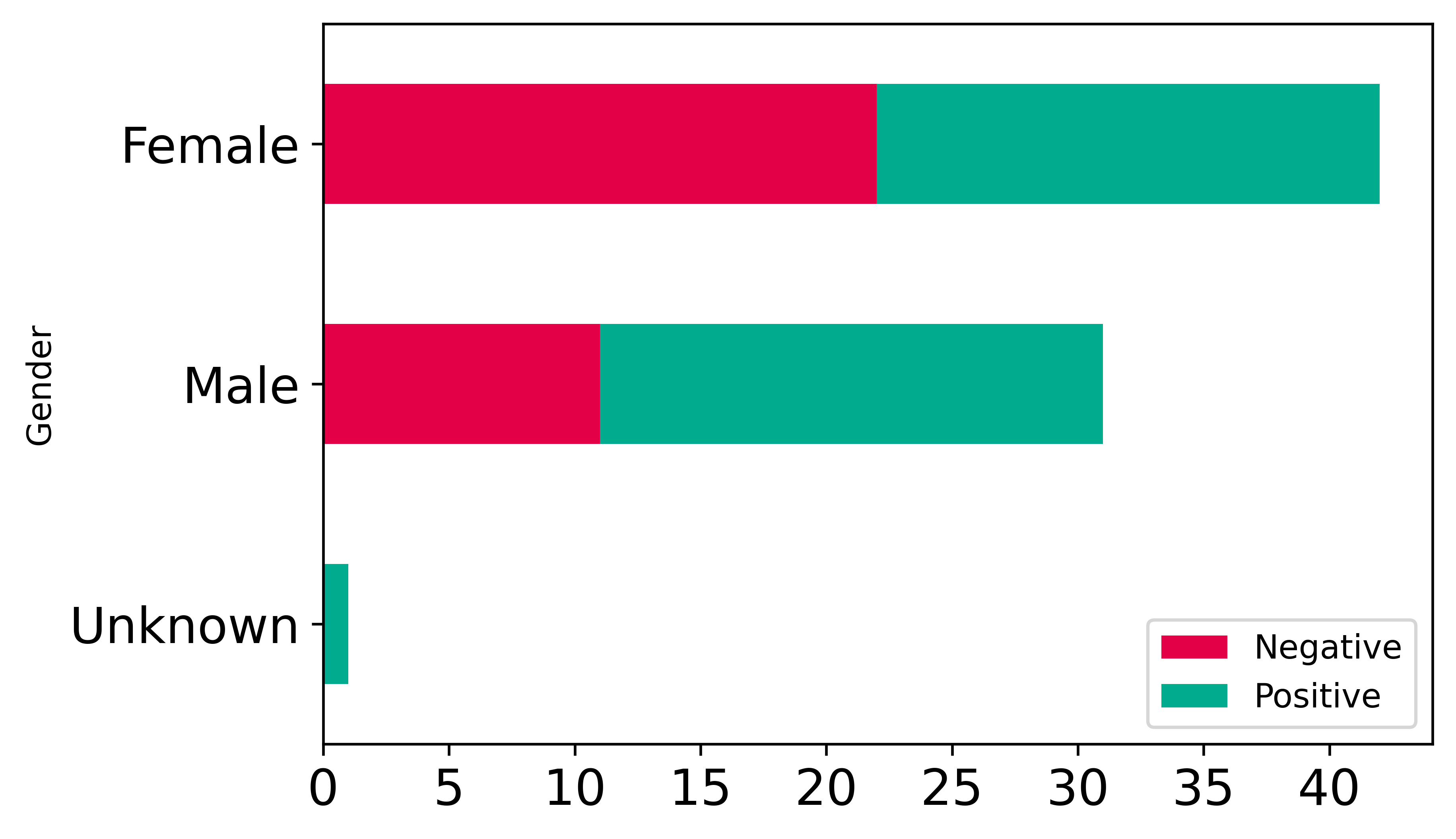} 
    \end{minipage} & 
    \begin{minipage}{.27\textwidth}
      \includegraphics[width=\linewidth]{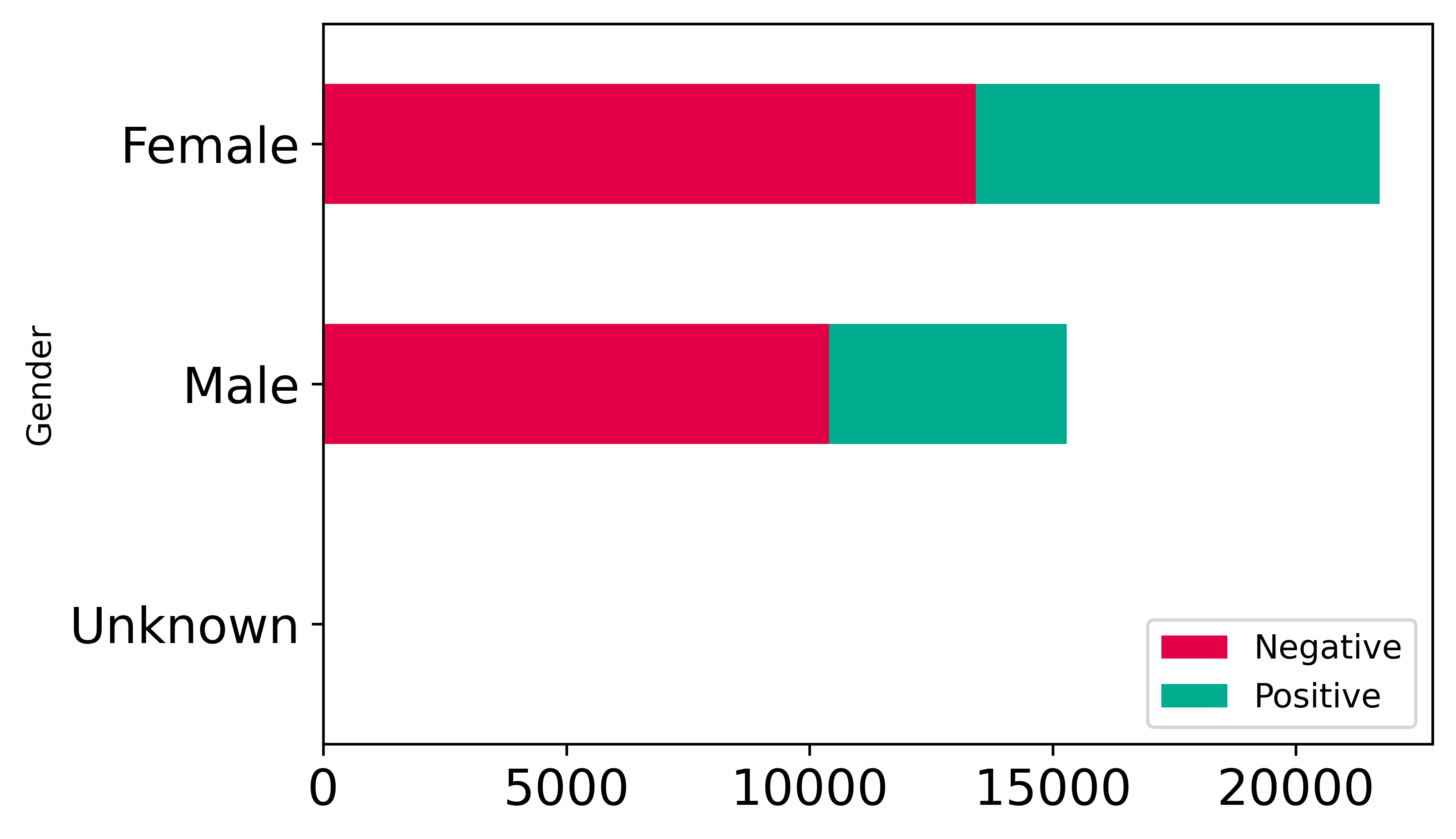}
    \end{minipage} \\ \hline 
    Test result & 
    \begin{minipage}{.27\textwidth}
      \includegraphics[width=\linewidth]{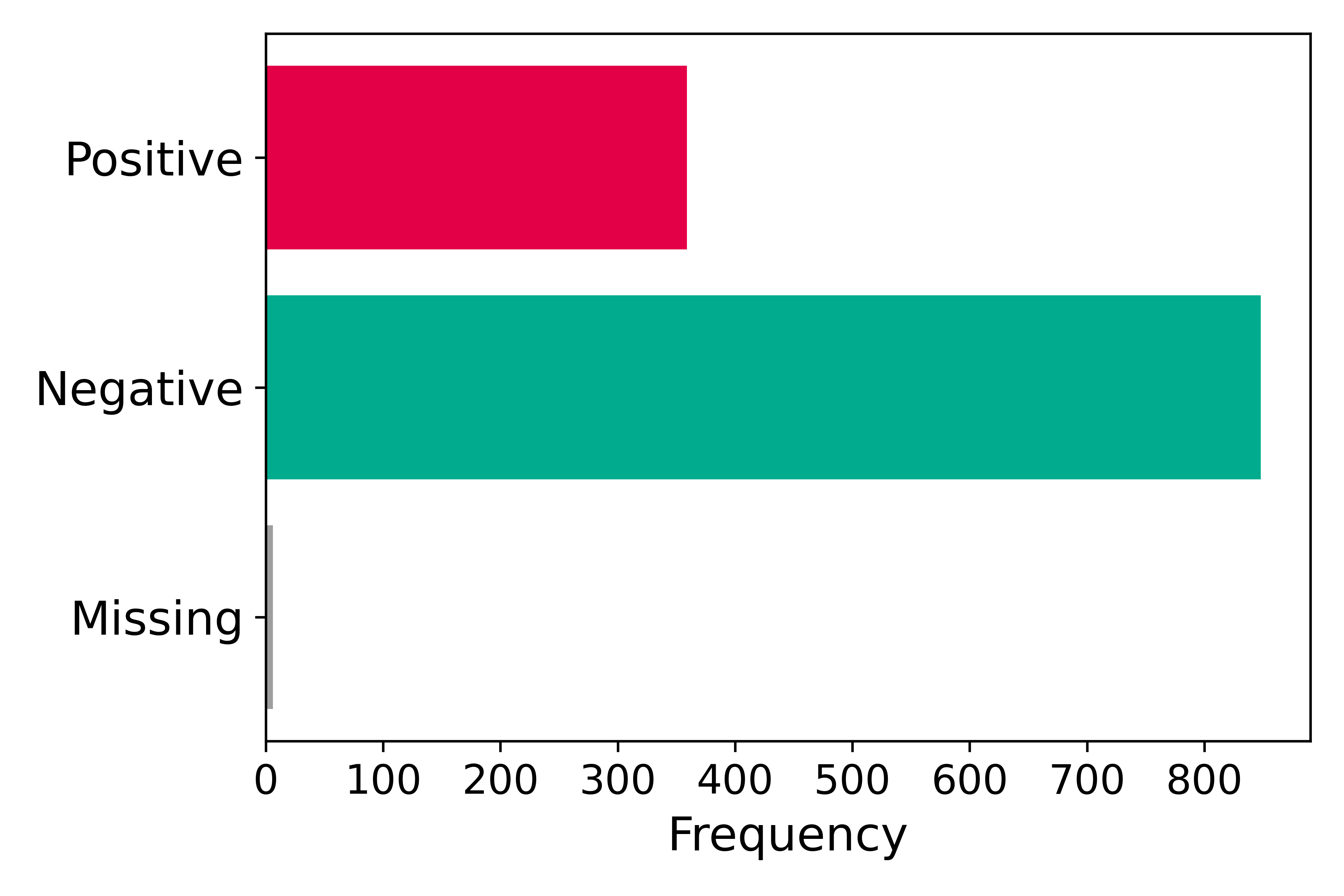}
    \end{minipage} & \begin{minipage}{.27\textwidth}
      \includegraphics[width=\linewidth]{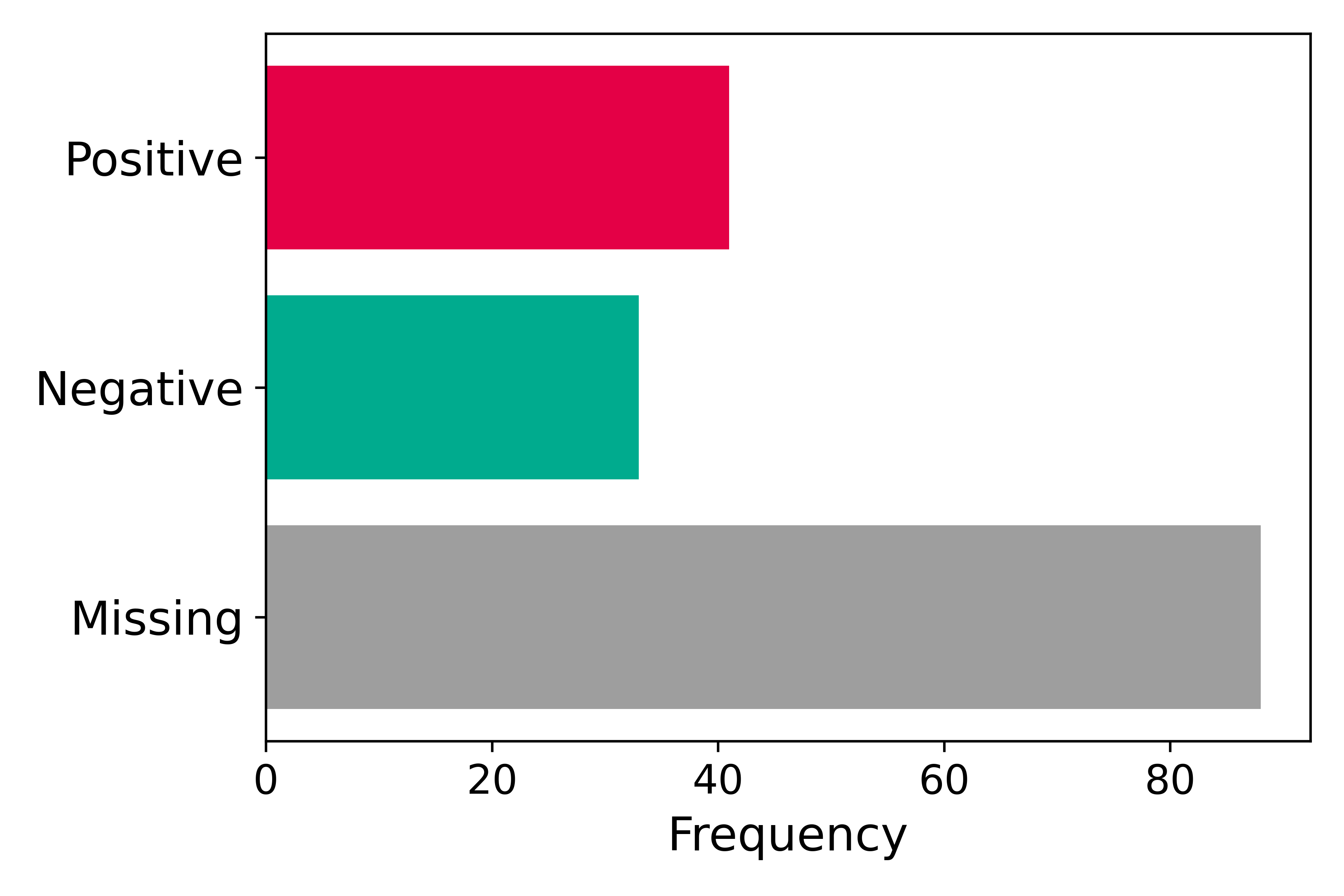} 
    \end{minipage} & 
    \begin{minipage}{.27\textwidth}
      \includegraphics[width=\linewidth]{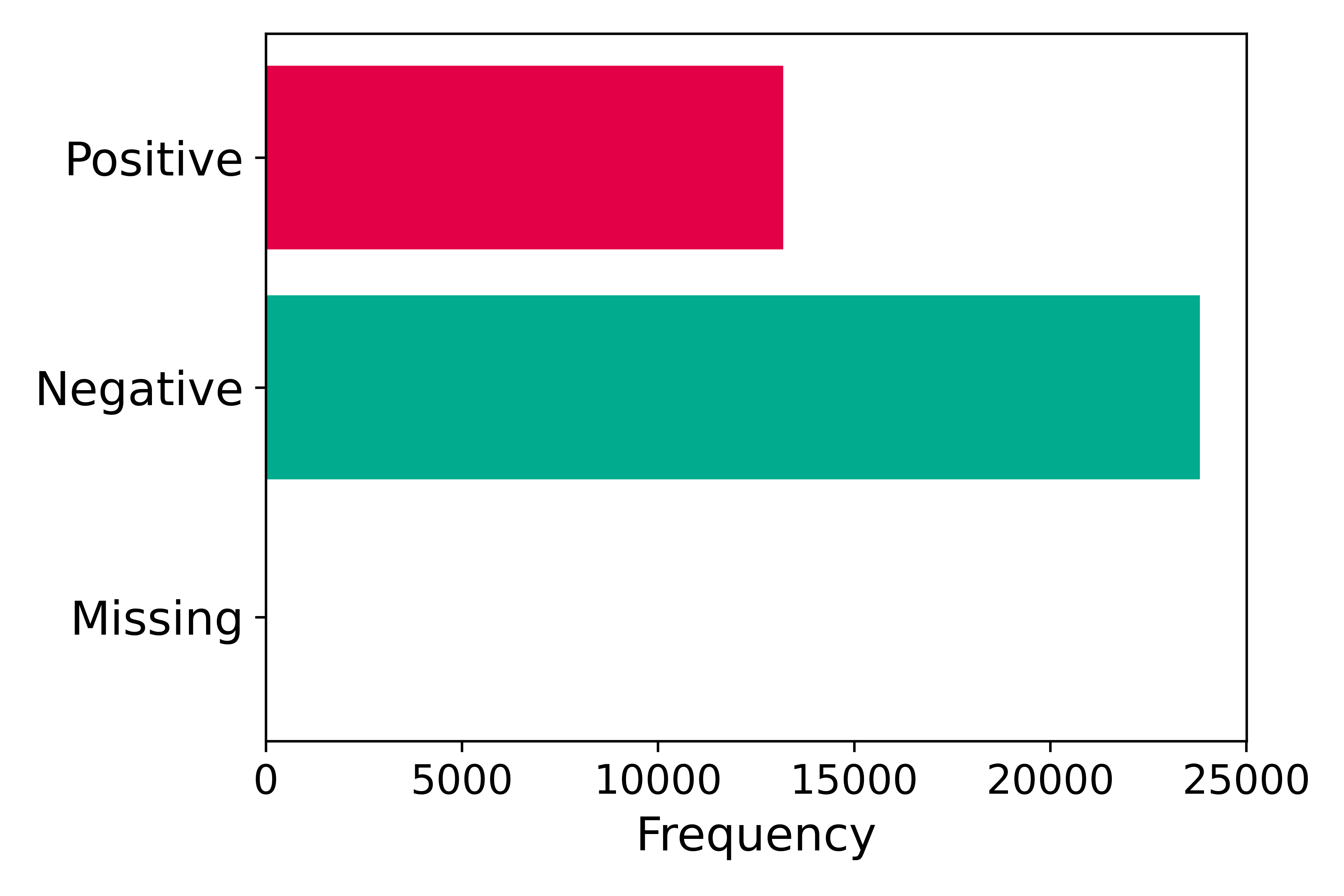}
    \end{minipage} \\ \hline
    Height by gender & 
    \begin{minipage}{.27\textwidth}
      \includegraphics[width=\linewidth]{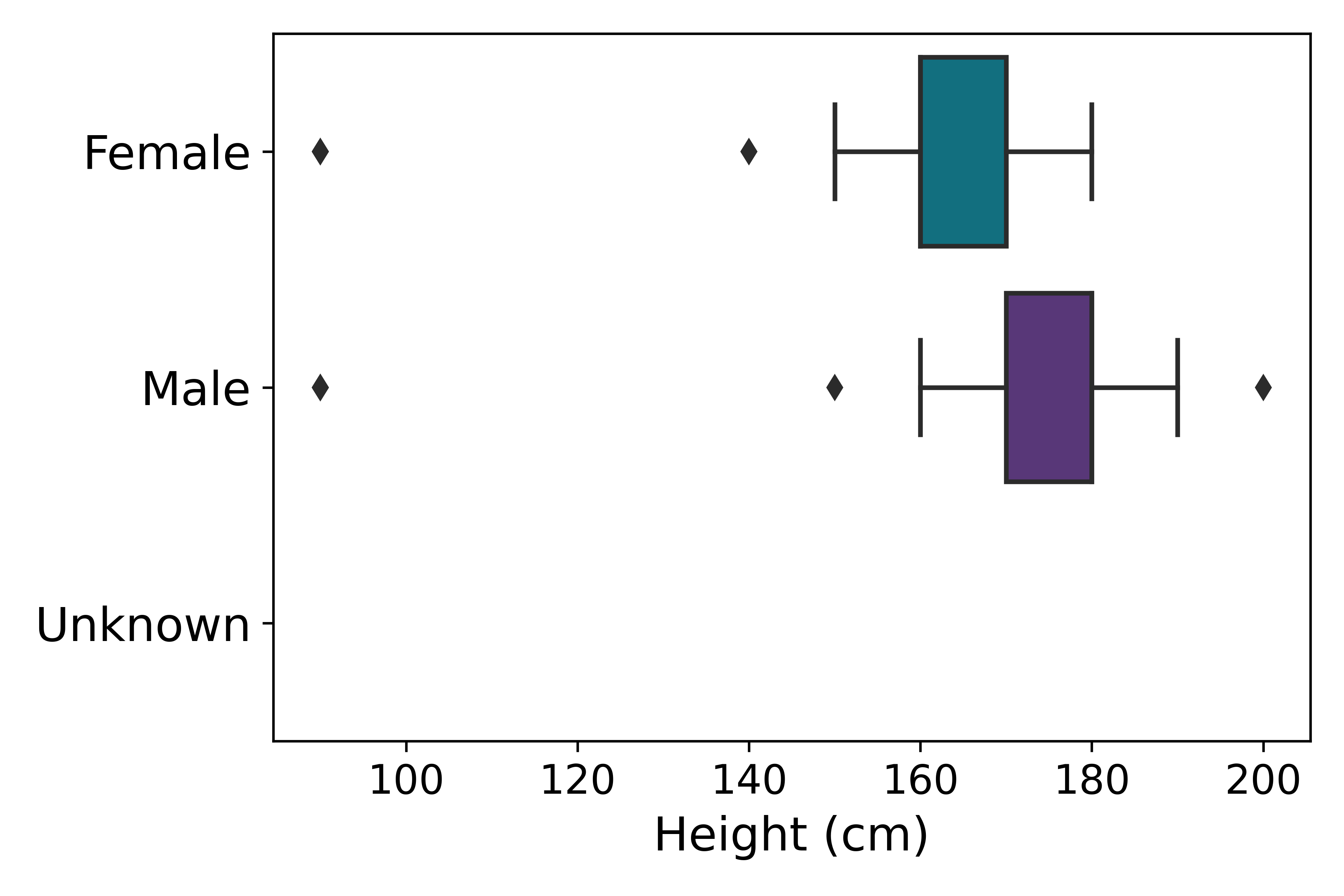} 
    \end{minipage} & \begin{minipage}{.27\textwidth}
      \includegraphics[width=\linewidth]{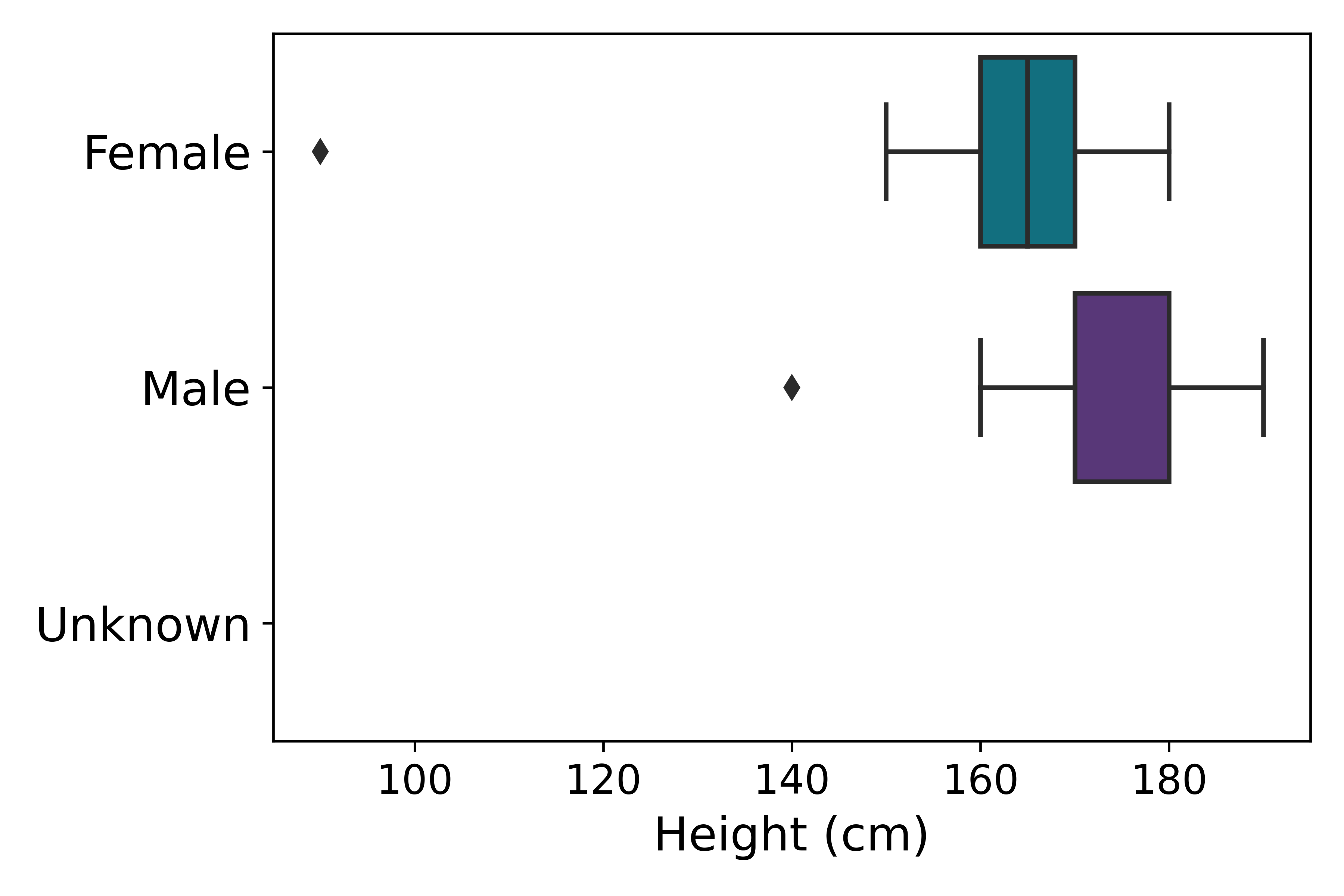} 
    \end{minipage} & 
    \begin{minipage}{.27\textwidth}
      \includegraphics[width=\linewidth]{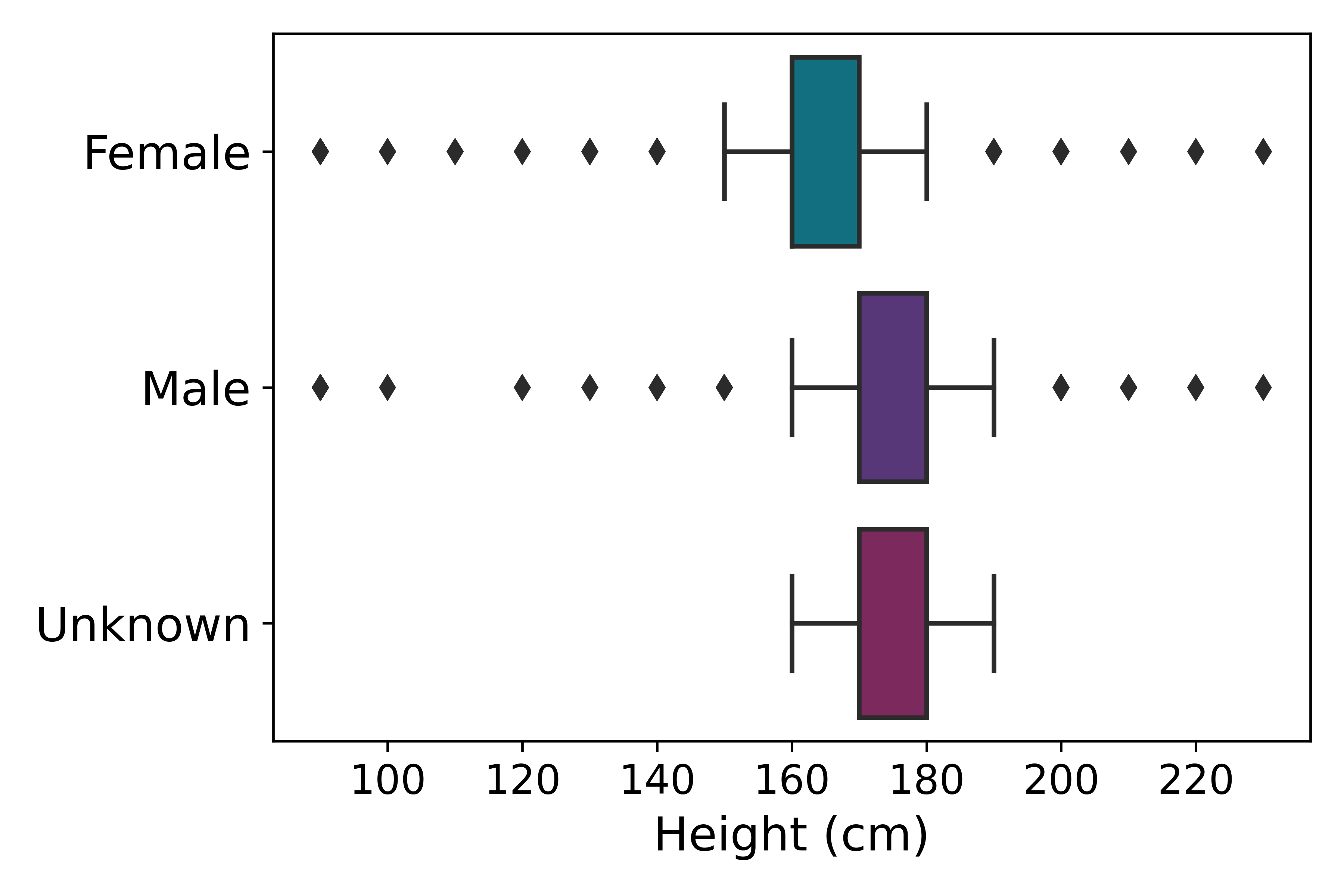} 
    \end{minipage} \\ \hline
    Age by test result & 
    \begin{minipage}{.27\textwidth}
      \includegraphics[width=\linewidth]{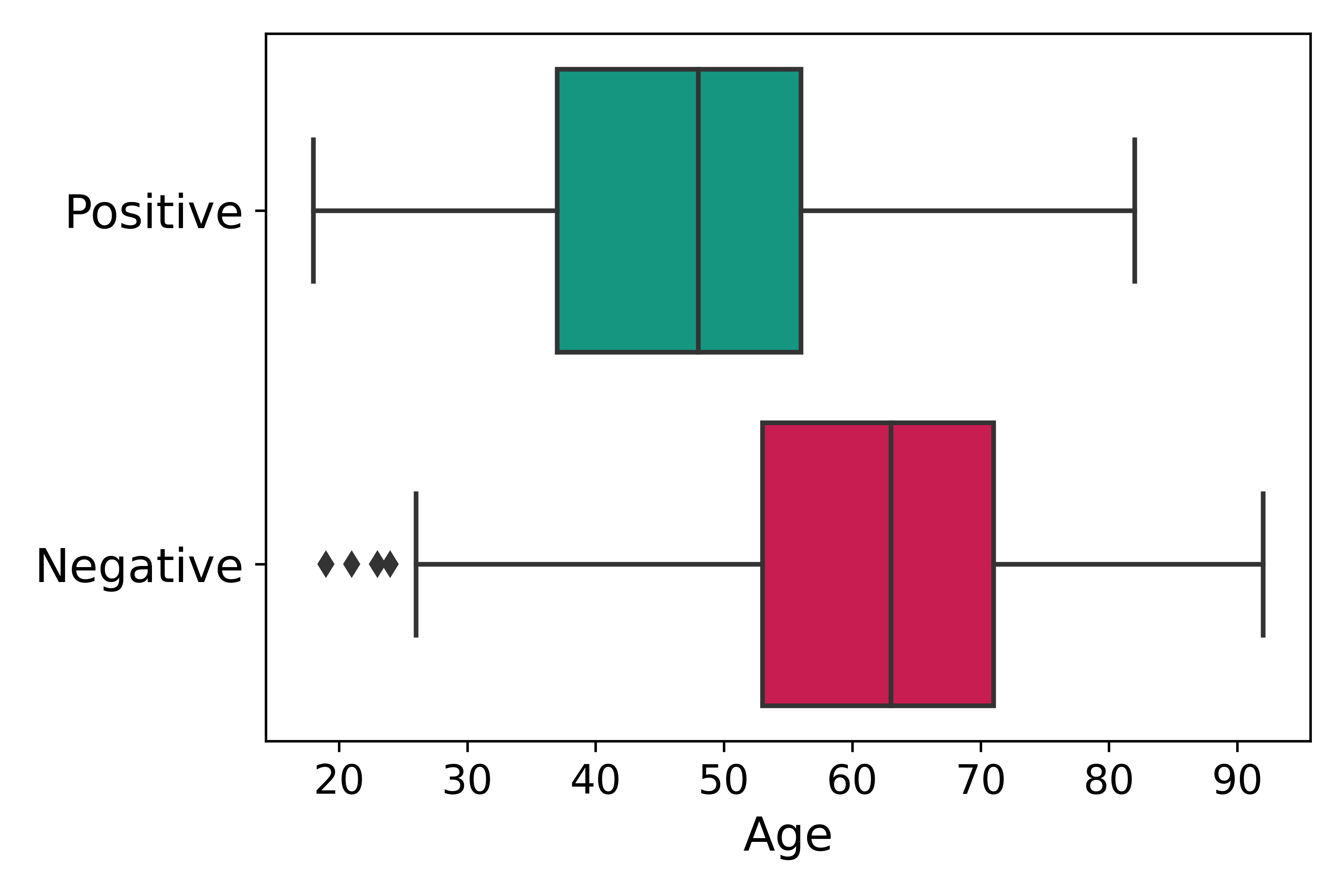} 
    \end{minipage} & \begin{minipage}{.27\textwidth}
      \includegraphics[width=\linewidth]{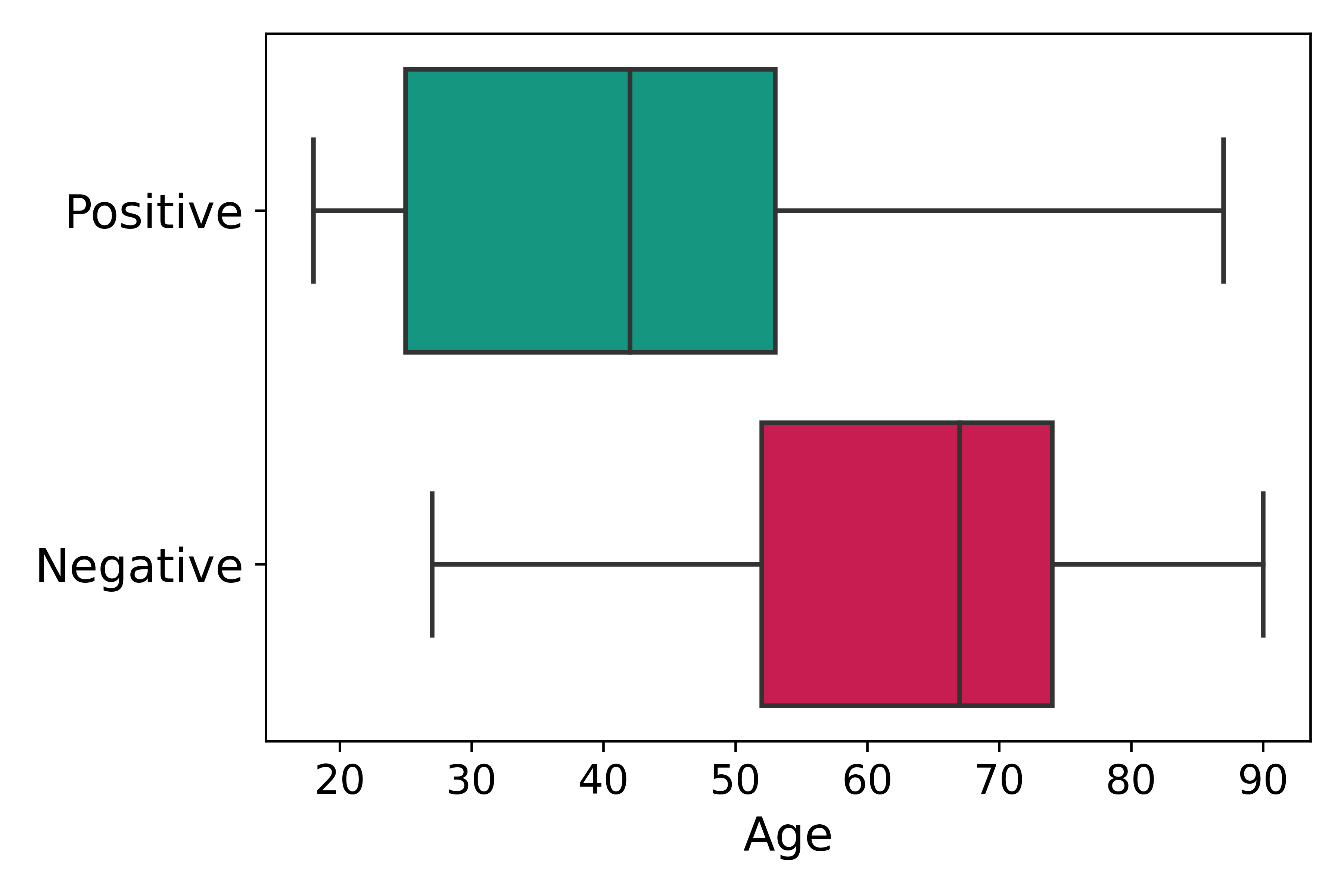} 
    \end{minipage} & 
    \begin{minipage}{.27\textwidth}
      \includegraphics[width=\linewidth]{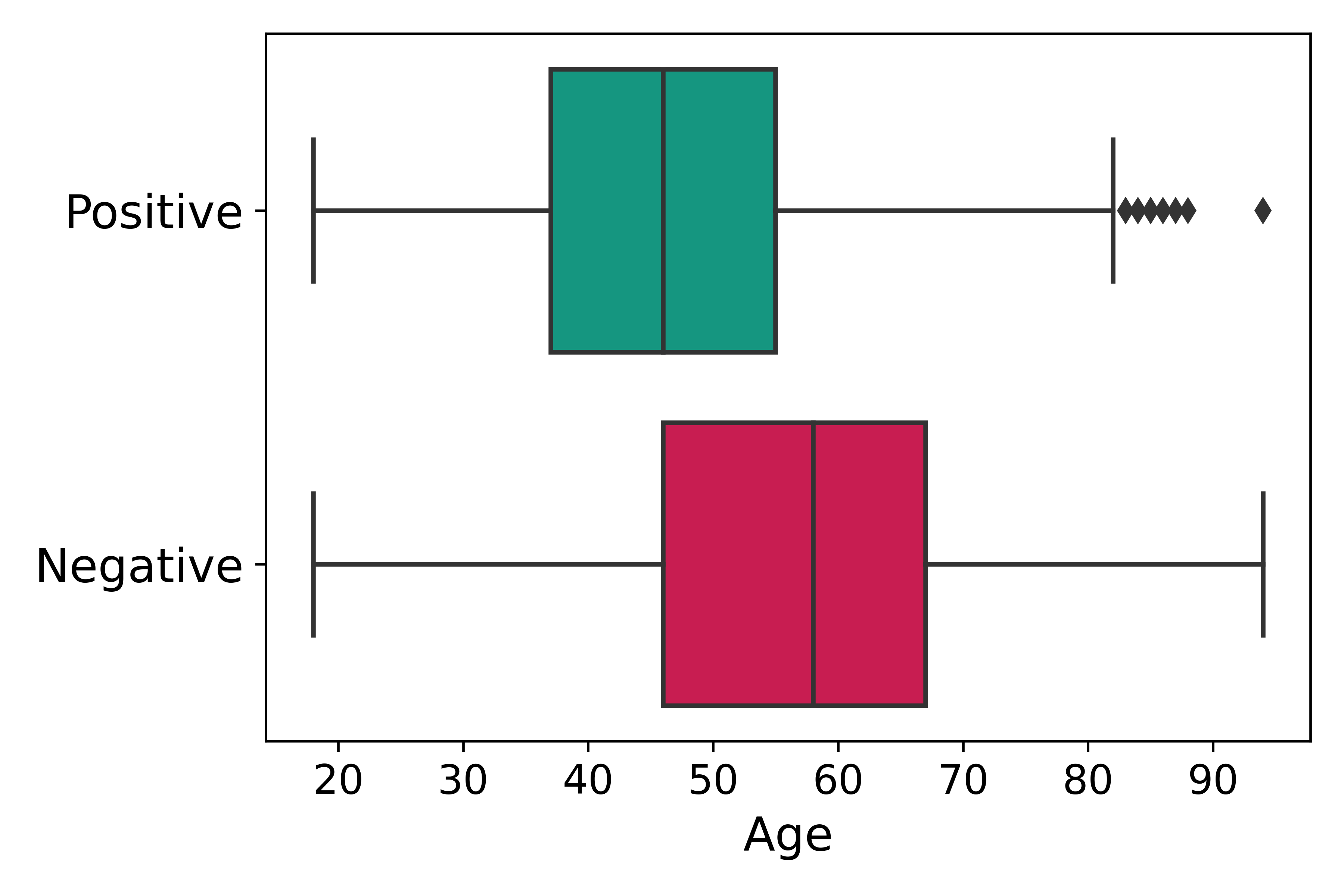} 
    \end{minipage} \\ \hline
    Symptoms & 
    \begin{minipage}{.27\textwidth}
      \includegraphics[width=\linewidth]{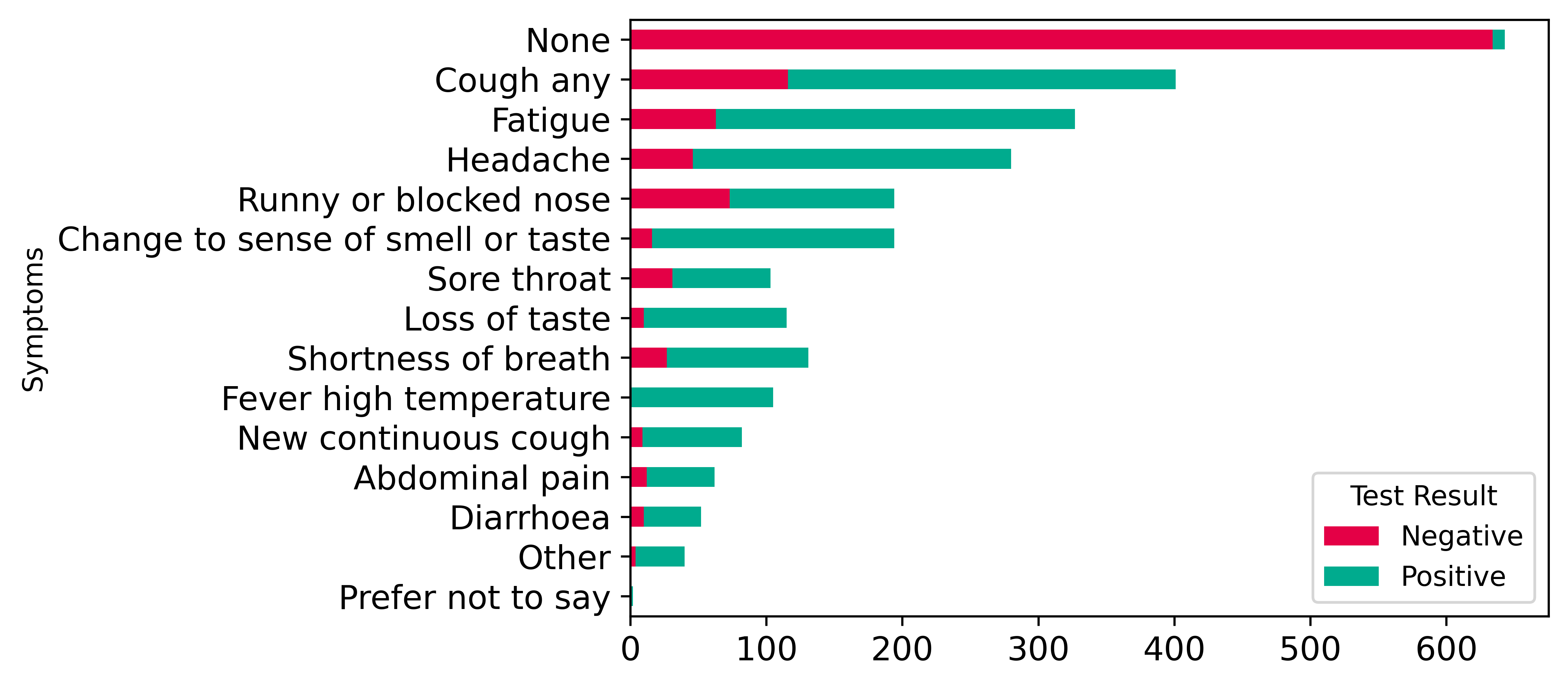} 
    \end{minipage} & \begin{minipage}{.27\textwidth}
      \includegraphics[width=\linewidth]{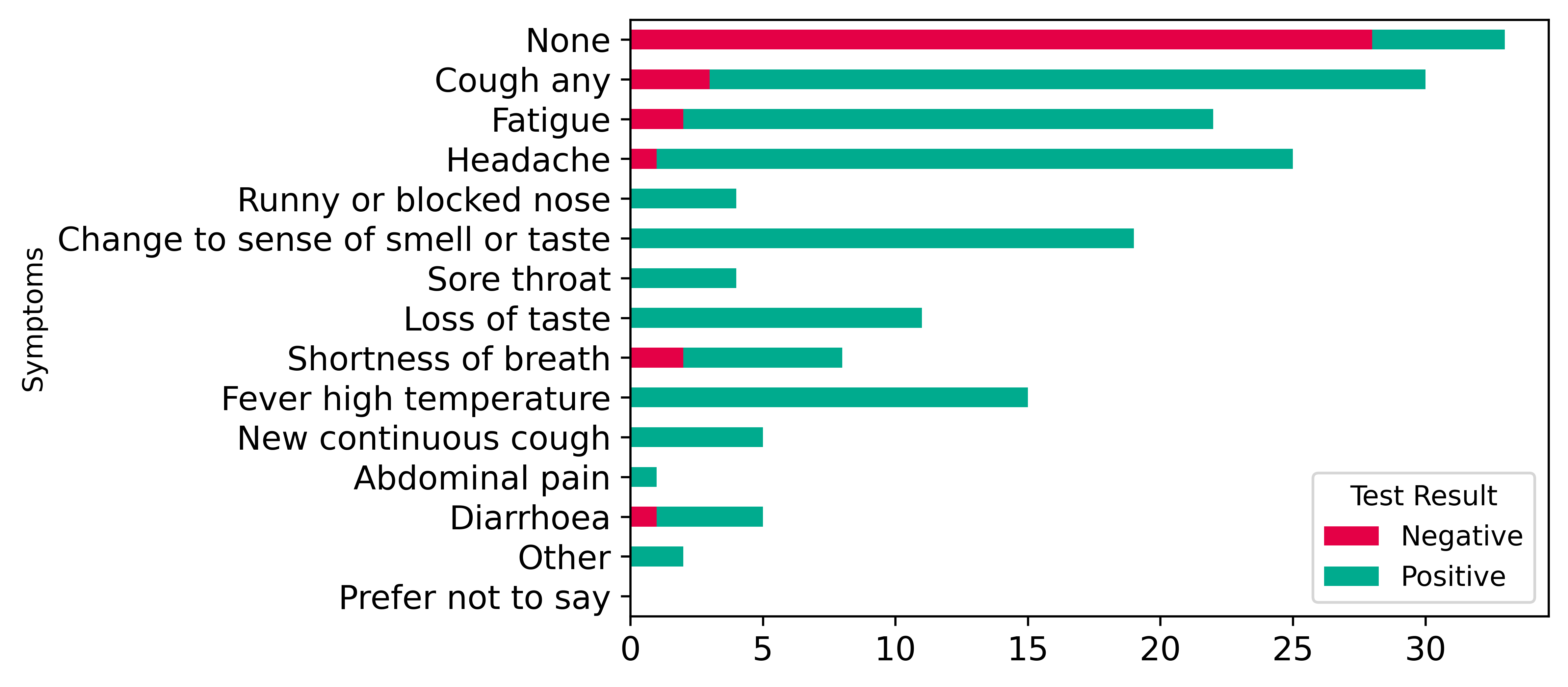} 
    \end{minipage} & 
    \begin{minipage}{.27\textwidth}
      \includegraphics[width=\linewidth]{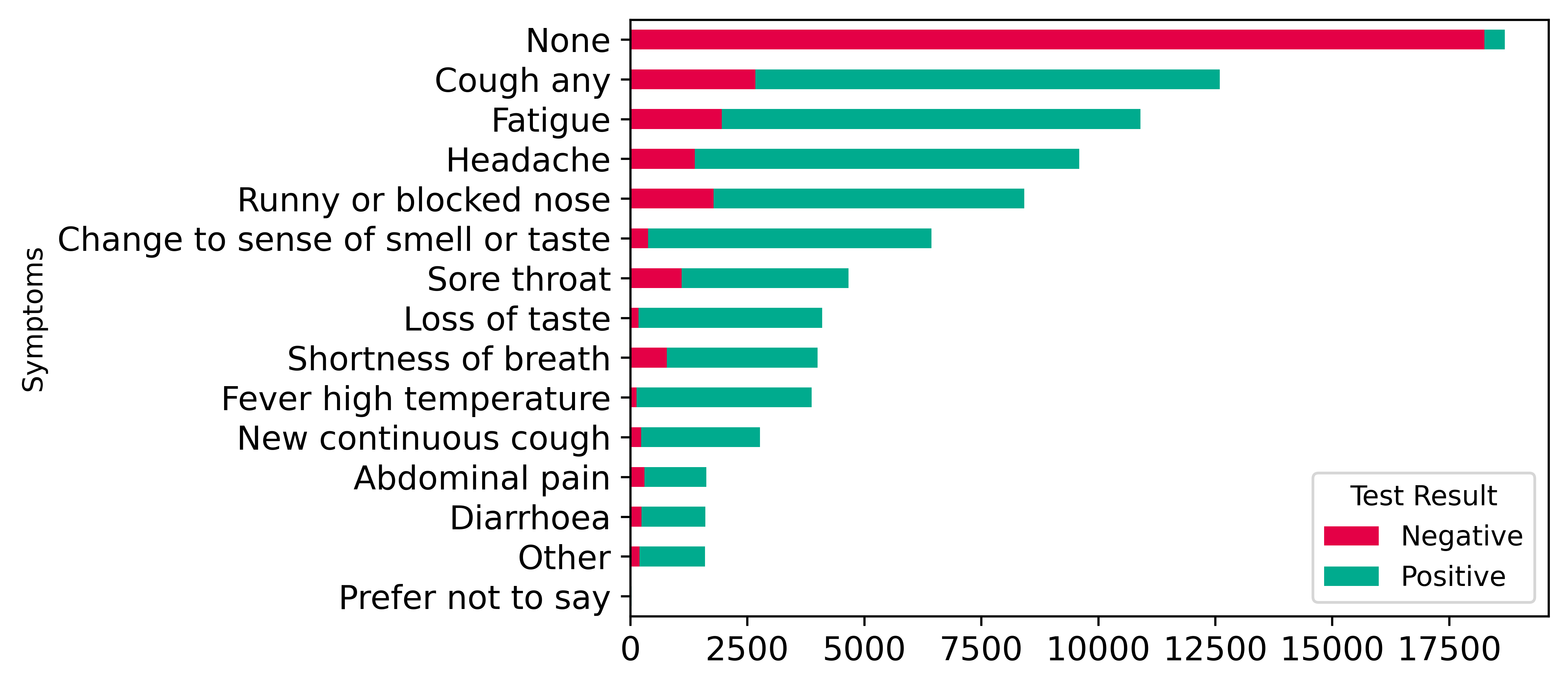} 
    \end{minipage} \\ \hline
    First Language & 
    \begin{minipage}{.27\textwidth}
      \includegraphics[width=\linewidth]{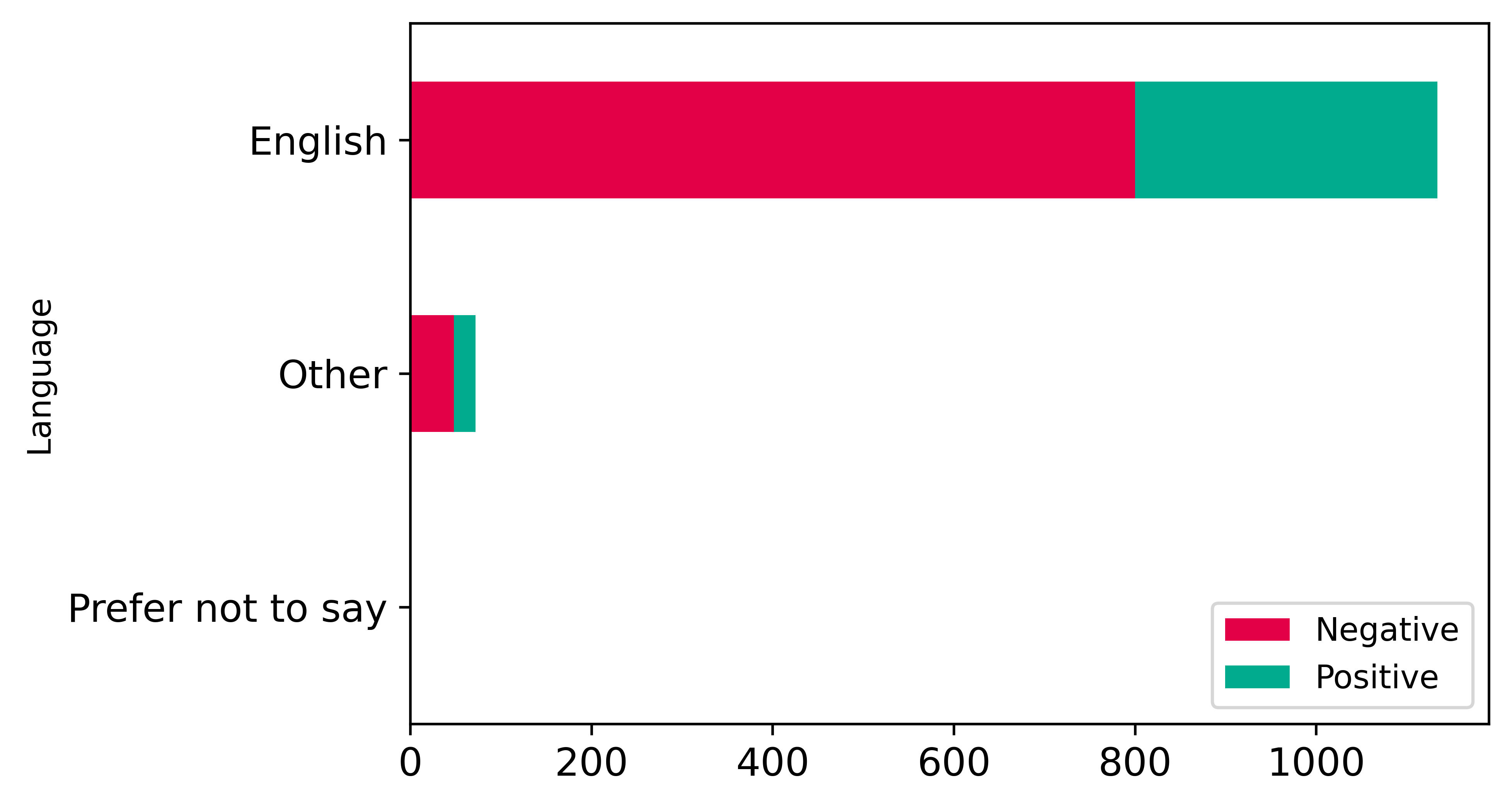} 
    \end{minipage} & \begin{minipage}{.27\textwidth}
      \includegraphics[width=\linewidth]{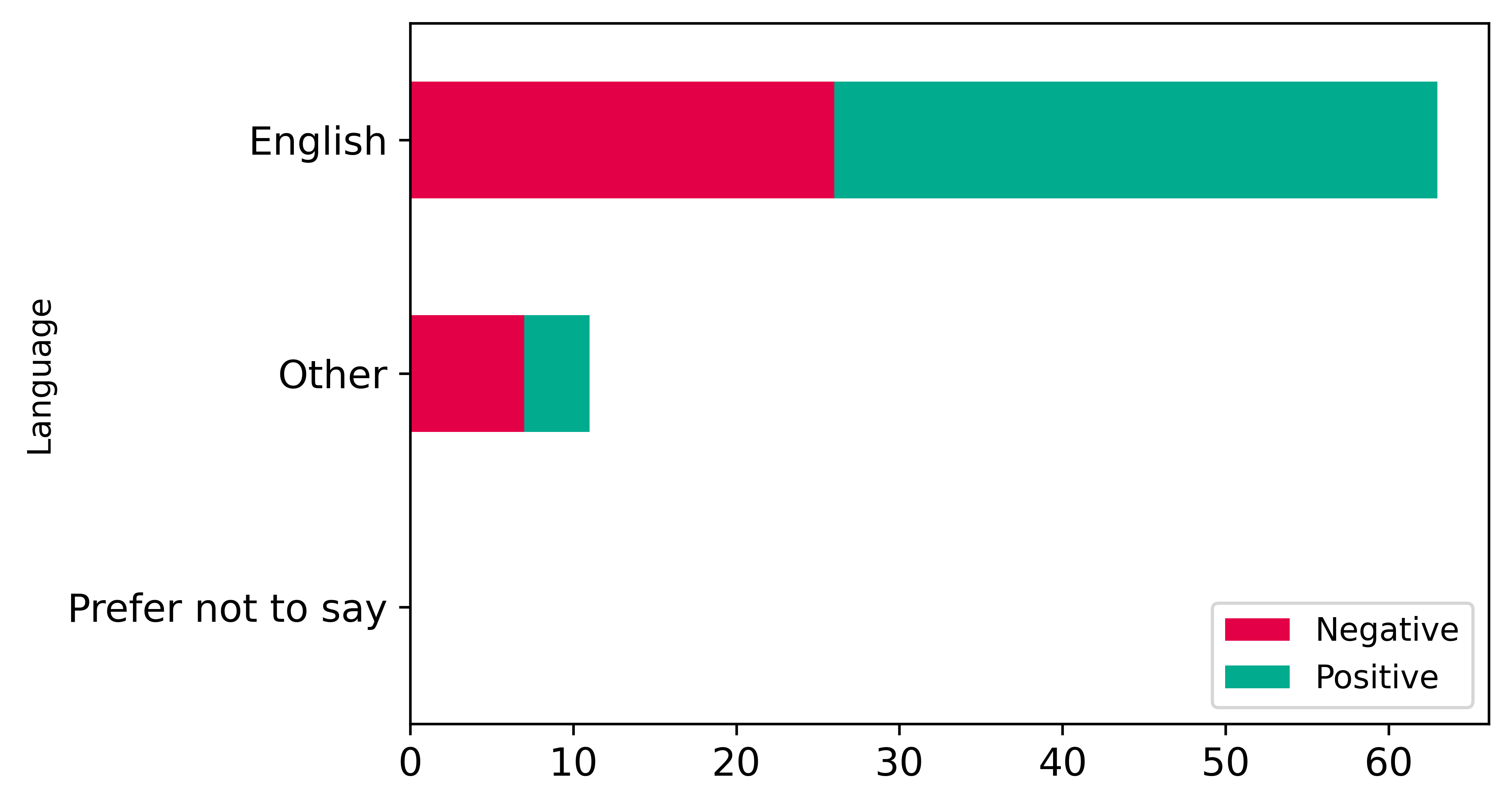} 
    \end{minipage} & 
    \begin{minipage}{.27\textwidth}
      \includegraphics[width=\linewidth]{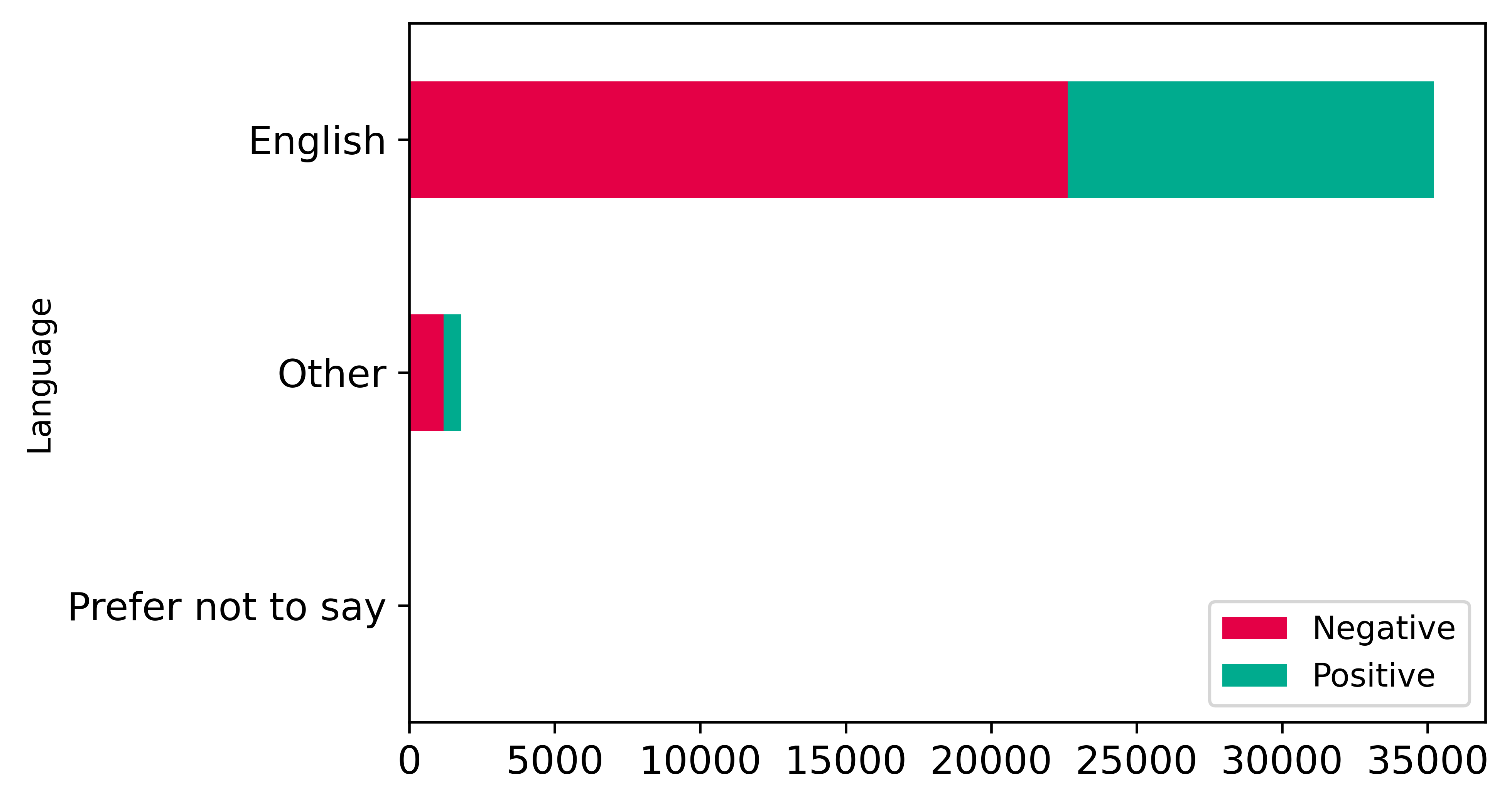} 
    \end{minipage} \\ \hline
    Weight by gender & 
    \begin{minipage}{.27\textwidth}
      \includegraphics[width=\linewidth]{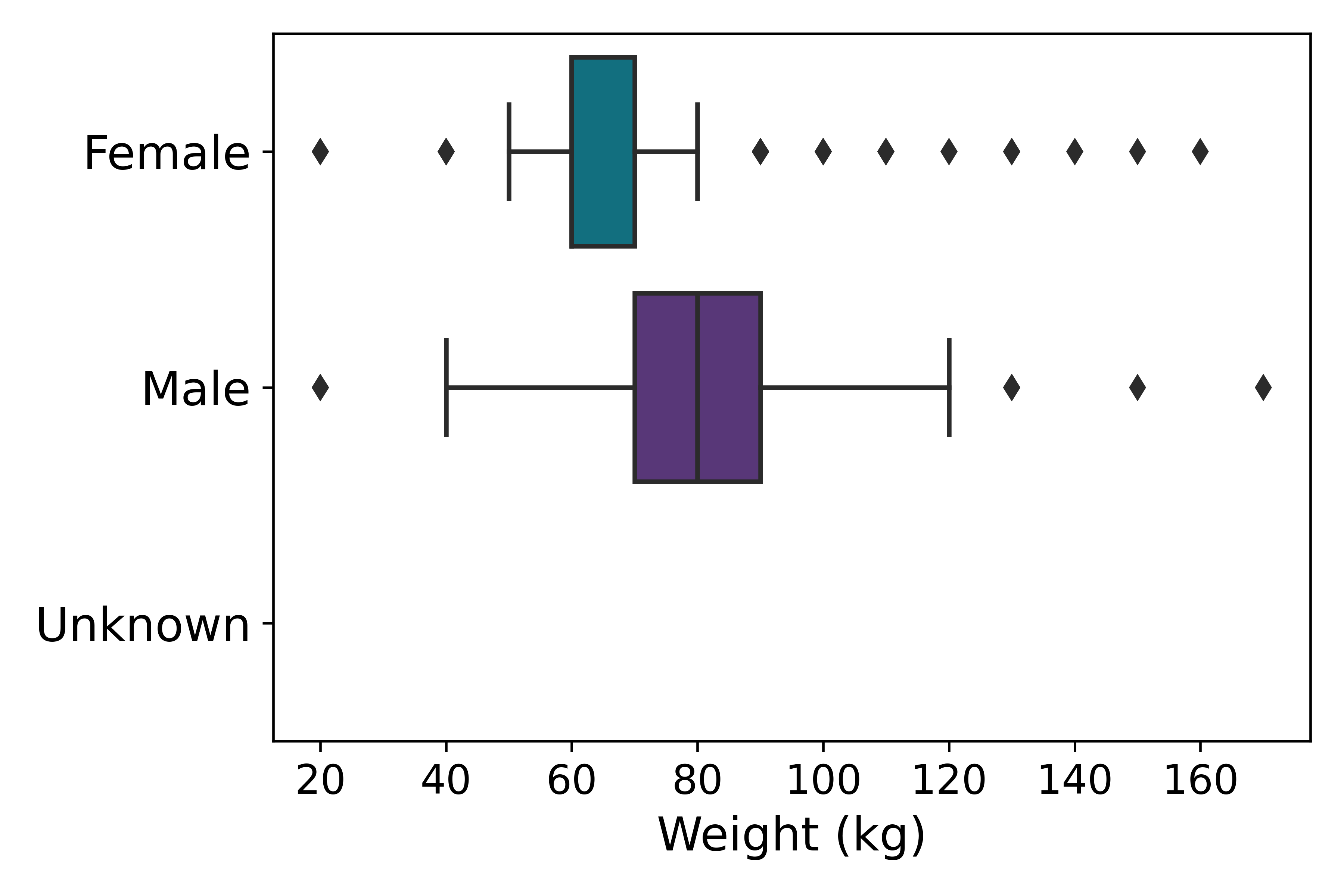} 
    \end{minipage} & \begin{minipage}{.27\textwidth}
      \includegraphics[width=\linewidth]{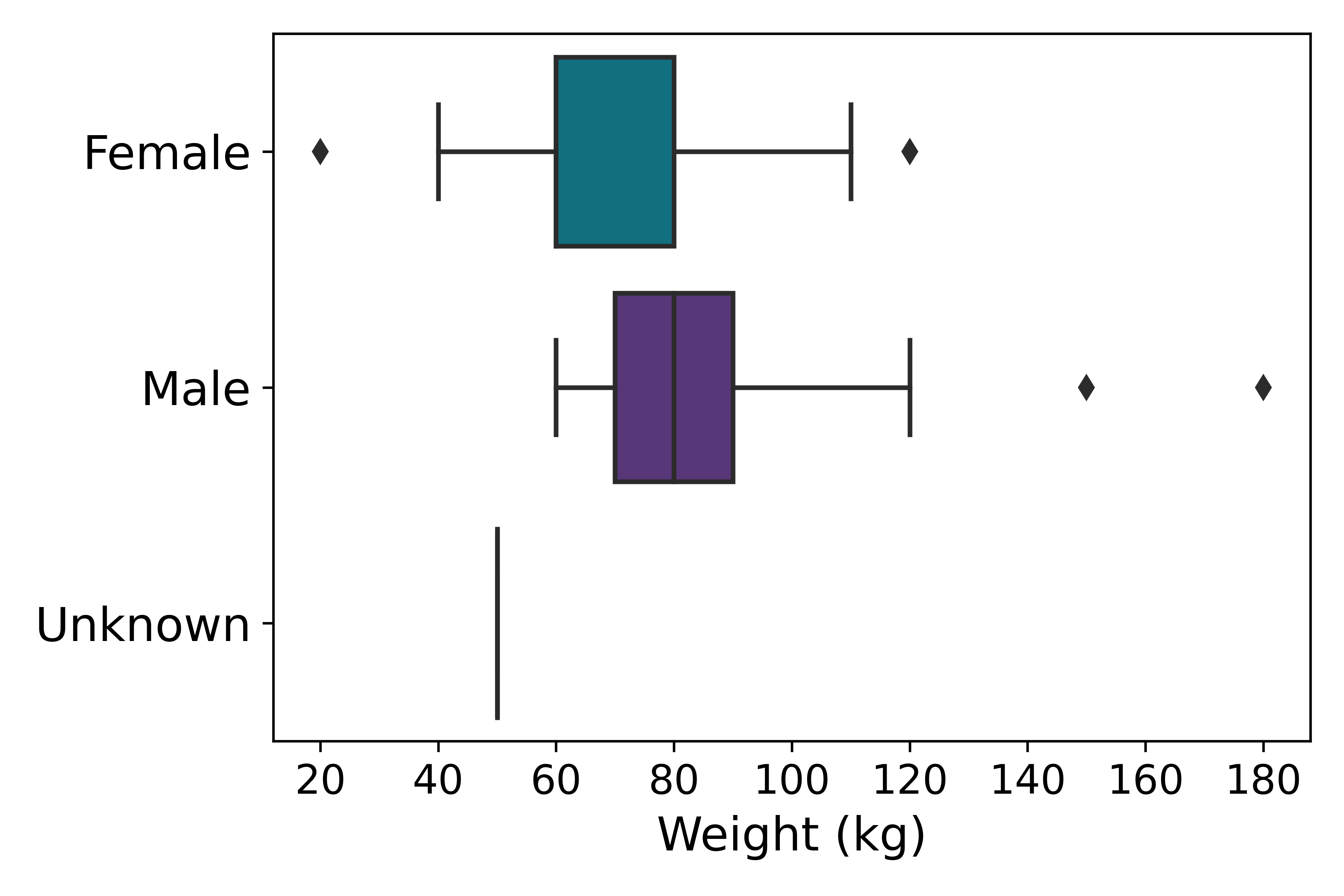} 
    \end{minipage} & 
     \begin{minipage}{.27\textwidth}
      \includegraphics[width=\linewidth]{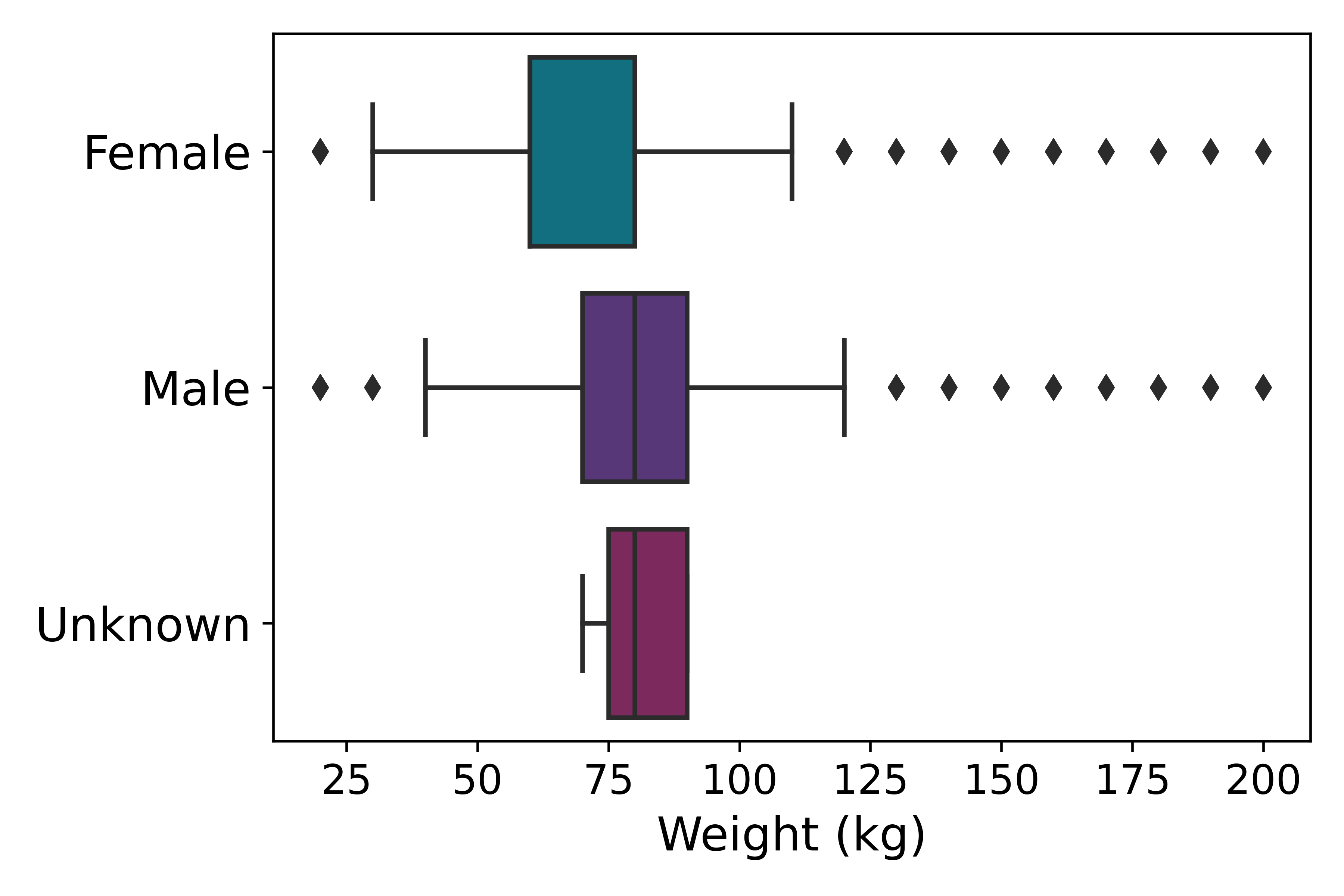} 
    \end{minipage} \\ 
    \hline
    \end{tabular}
    \caption{
\label{fig:metadatabreakdown} Breakdown of the available meta-data for the portions of the dataset with missing audio features, missing meta-data and complete data, i.e., the final dataset used in the following analysis. We cannot see any problematic imbalances that may suggest that audio-feature or meta-data are not missing at random. }
\end{figure}

Participants in the ``Speak up and help beat coronavirus'' study were asked to complete an online survey on a smartphone, tablet, or computer. Health and demographic information and audio recordings made via the microphone of the participant's device were collected. These data were linked with the participants' COVID-19 PCR test results and associated information via a patient identifier, as described in \cite{ciab_data}. In the following, we are going to refer to submissions linked to a positive or negative COVID-19 PCR test result as COVID-19 positive and COVID-19 negative submissions. Also, while technically the PCR test is an indicator of SARS-CoV-2 infection status, in this paper we are going to refer to it as COVID-19 infection status (positive or negative), since this terminology has become widespread in the public health communication. Each submission consists of four audio modalities
(one and three successive forced coughs, three successive exhalation sounds, and a full sentence read from text), and metadata including individual-level information (age, gender, ethnicity, local authority, existing respiratory health conditions, smoker status, first language, height, weight), PCR test result (SARS-CoV-2+ or SARS-CoV-2-), self-reported symptoms as listed in Figure \ref{fig:metadatabreakdown}, date of the PCR test, date of the audio submission and recruitment source (NHS Test and Trace contact or REACT-1 survey). PCR cycle threshold and resulting viral load estimation, as well as vaccination status information, is available for some participants. Further details on the recruitment procedures and the dataset are available in \cite{ciab_data}.

For the analysis described in this paper, data was then removed for participants if they did not satisfy all the following conditions:
\begin{itemize}
    \item Participant aged 18 or over.
    \item Test results obtained via PCR.
    \item Audio recordings submitted between the test result and 10 days after the test result.
    \item Test performed by a lab not under investigation by UK Health Security Agency for inaccurate results.
    \item No discrepancy in recording of symptoms (both symptoms and no symptoms selected)
\end{itemize}
Submissions with missing data, either in terms of missing one or more of the audio submissions or missing some of the meta-data, were removed. The resulting dataset without missing data was comprised of 37018 submissions.

Figure \ref{fig:metadatabreakdown} shows a breakdown of the meta-data for the missing data submissions and for the final dataset (disclosure control measures have been implemented where categories with less than 5 participants have been shown, see \cite{ciab_data} for more details). Since this does not highlight any systematic patterns and the overall number of instances with missing data is modest, submissions with missing data were removed from the analysis. 

\section{Analysis of Meta-data}

We carried out a descriptive analysis of the meta-data to explore potential biases in the data collection and to design a test set which aims to provide a reliable estimate of the performance in the general population.

\begin{figure}
    \centering
         \includegraphics[width=\textwidth]{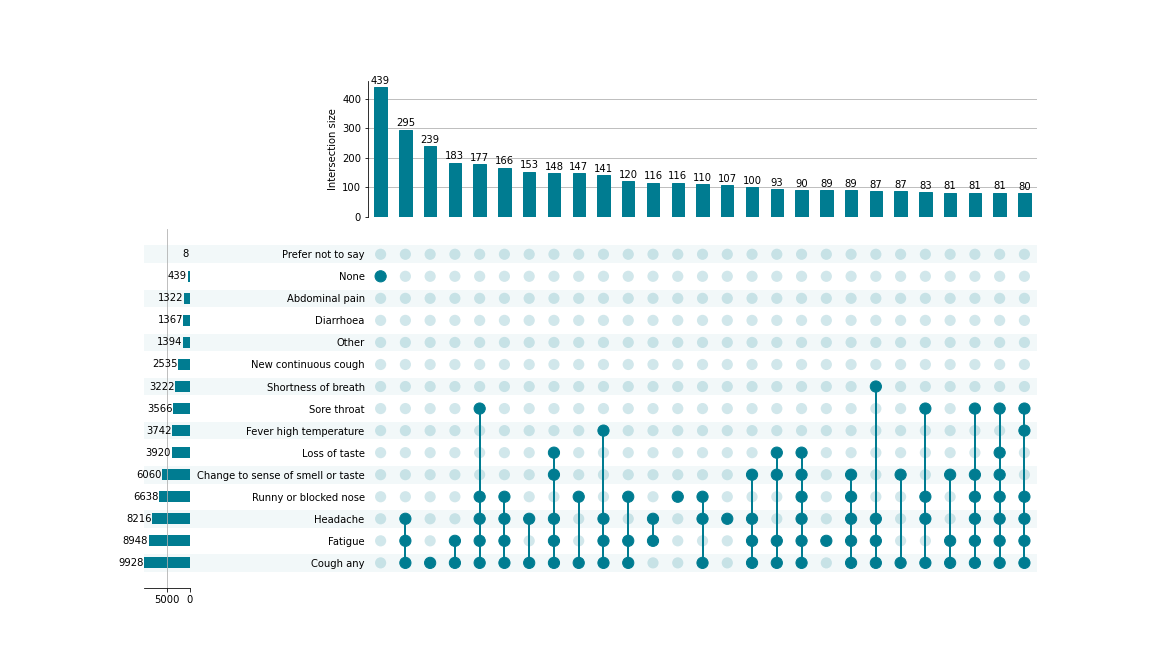}\\
         \includegraphics[width=\textwidth]{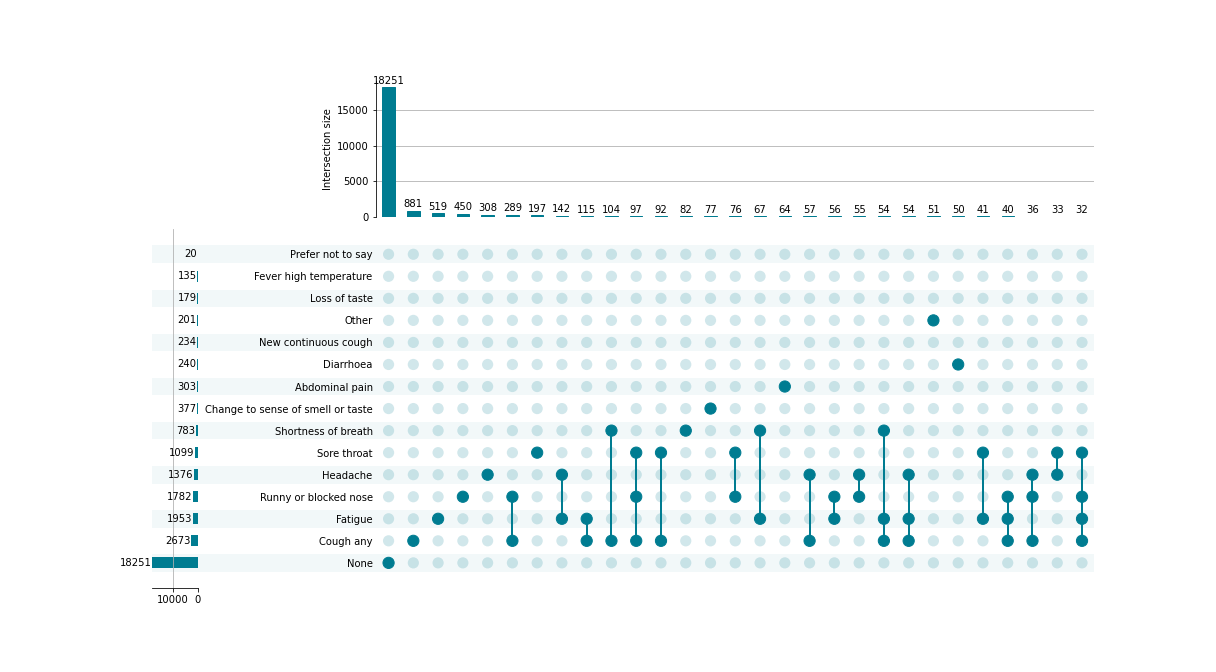}
    \caption{Distributions of self-reported symptoms for the submissions with positive PCR results (top) and negative PCR results (bottom). Each symptom is listed on the left, ordered by the frequency of appearance of each symptom. The combinations of each pair of symptoms is represented in the vertical stripes within each figure, ordered by the frequency of the symptom combinations shown at the top of each plot. Frequency cut offs of 80 and 30 
    for symptom combinations are used for the positive PCR and negative PCR plots respectively.}
    \label{fig:symptoms}
\end{figure}

\paragraph{Self-reported symptoms.} The distribution of the self-reported symptoms varies dramatically between the submissions with COVID-19 positive status and those with COVID-19 negative status (see Figure \ref{fig:symptoms}), thus raising the question of potential confounding between symptoms and COVID-19 status. We can see that for COVID-19 negative submissions, there is a dominance of no-symptoms, and very few symptomatic negative cases. For COVID-19 positive cases, there are only 446 asymptomatic submissions – a key group for understanding the efficacy of models in detecting COVID-19 infection status. It is also worth noting the dominance of the cough symptom as the most common symptom in both positive and negative cases.

\begin{figure}[h]
    \centering
    \includegraphics[width=0.55\textwidth]{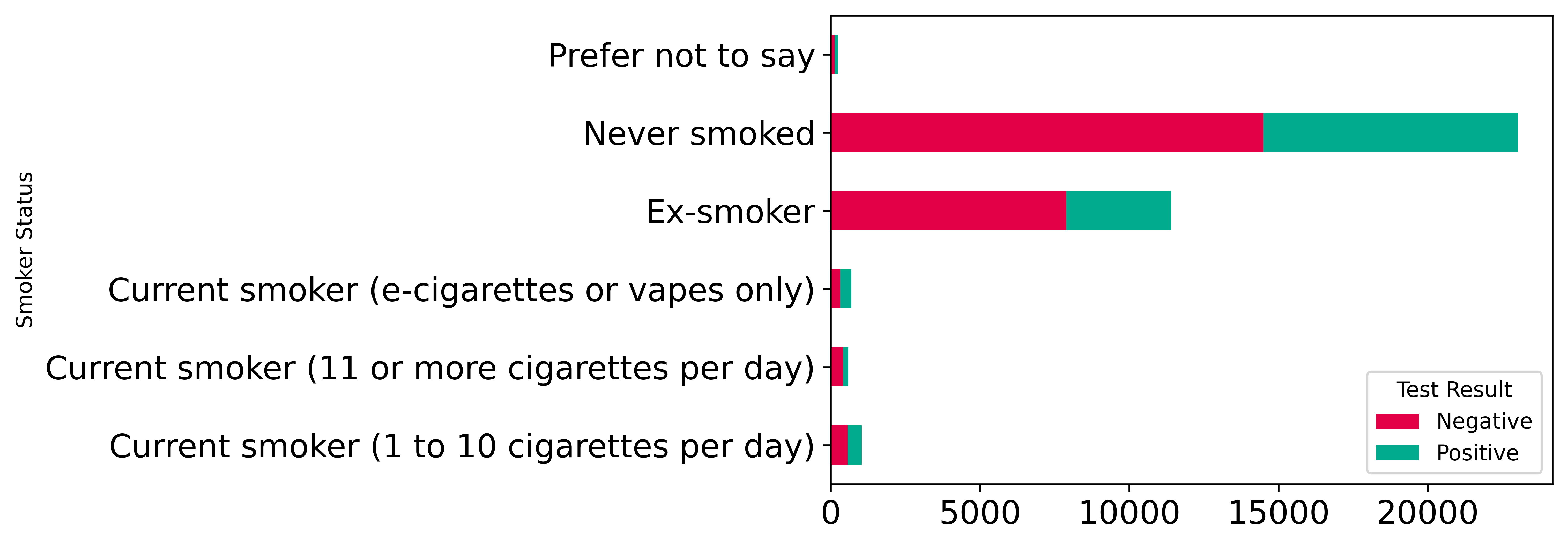}
    \includegraphics[width=0.4\textwidth]{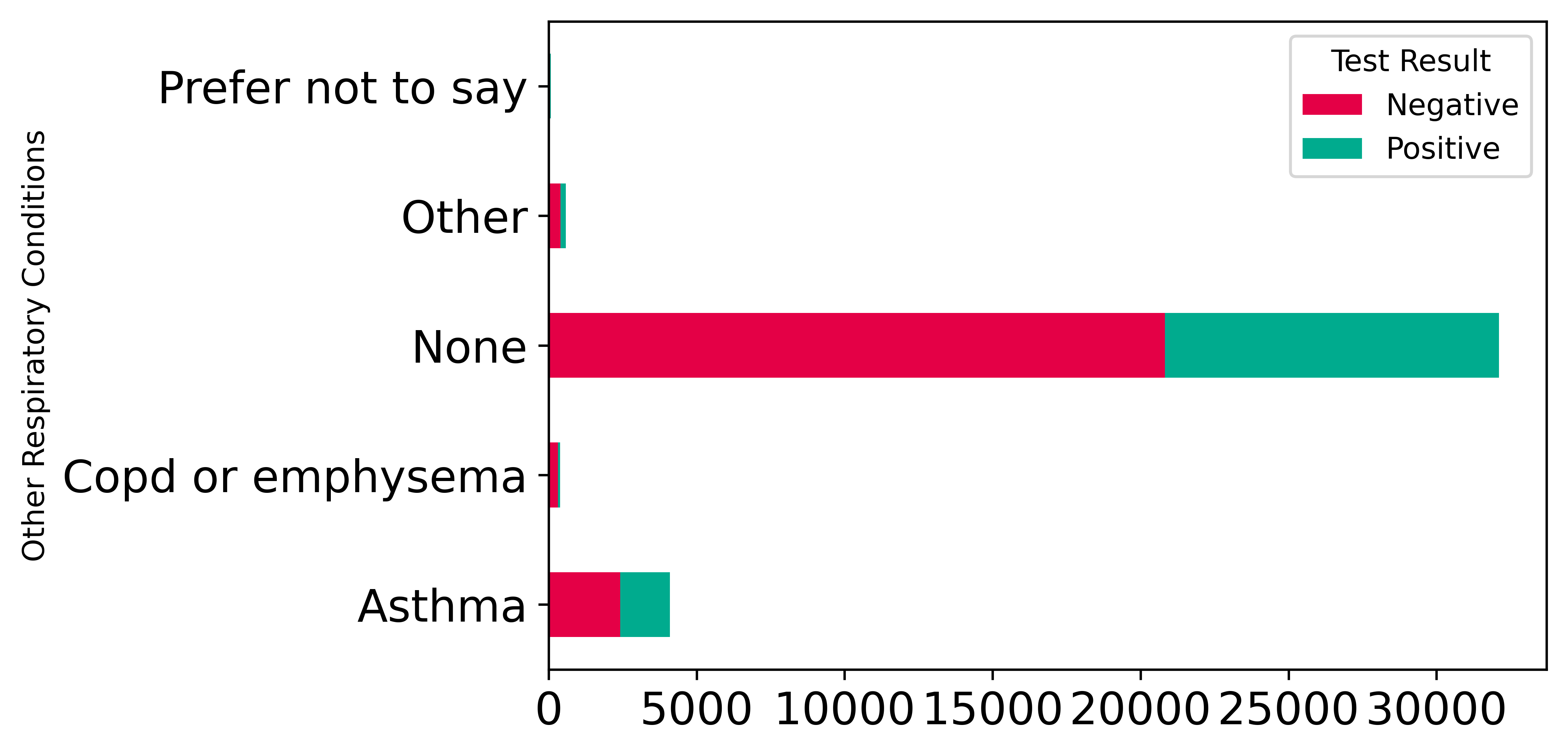}
    \caption{Left: Barplots of absolute frequencies of smoker status in the final dataset by COVID-19 status. Right: Barplots of absolute frequencies of respiratory conditions in the final dataset by COVID-19 status. }
    \label{fig:covid_by_smoker}
\end{figure}

\paragraph{Other respiratory conditions and smoker status.}
Information related to other respiratory conditions and smoker status was captured in the data collection due to possible effects on the nature of someone’s cough, e.g., if a person is asthmatic, or an ex-smoker. Stacked bar plots and cross tabulations of these are included in Figure \ref{fig:covid_by_smoker}. Neither of these appear to be associated with COVID-19 infection status in our dataset.

\begin{figure}[h]
    \centering
    \includegraphics[scale=0.47]{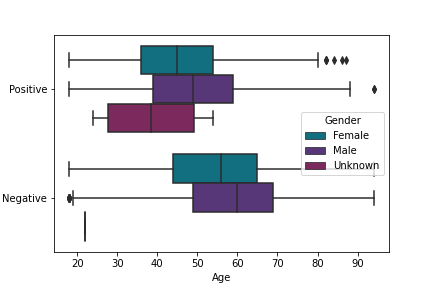}
    \includegraphics[scale=0.47]{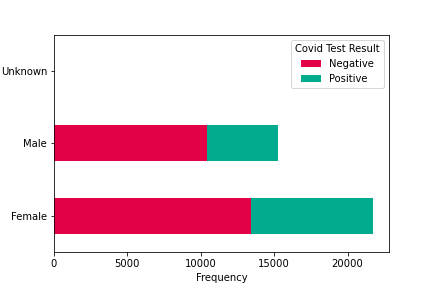}
    \caption{Left: Distribution of Age by COVID-19 status. Right: Frequency of Gender by COVID-19 status.}
    \label{fig:covid_by_age_gend}
\end{figure}

\paragraph{Age, Gender, Height and Weight.} 
There is some evidence supporting the hypothesis that age might act as a confounder of COVID-19 status in the dataset, with a dominance of older negatives and younger positives. This could be caused by any number of behavioural, societal and economic variations between younger and older people, as well as increased vaccination levels in the older population – this tendency is clearly seen in Figure \ref{fig:covid_by_age_gend}. There is a difference of 12 years between the median ages of those who tested positive and those who tested negative. Note that the median age for those testing positive without symptoms was 47 years.
There does not appear to be any confounding between gender and COVID-19 status (see Figure \ref{fig:covid_by_age_gend}), although there are 6,426 more females than males.

\begin{figure}[h]
    \centering
    \includegraphics[scale=0.45]{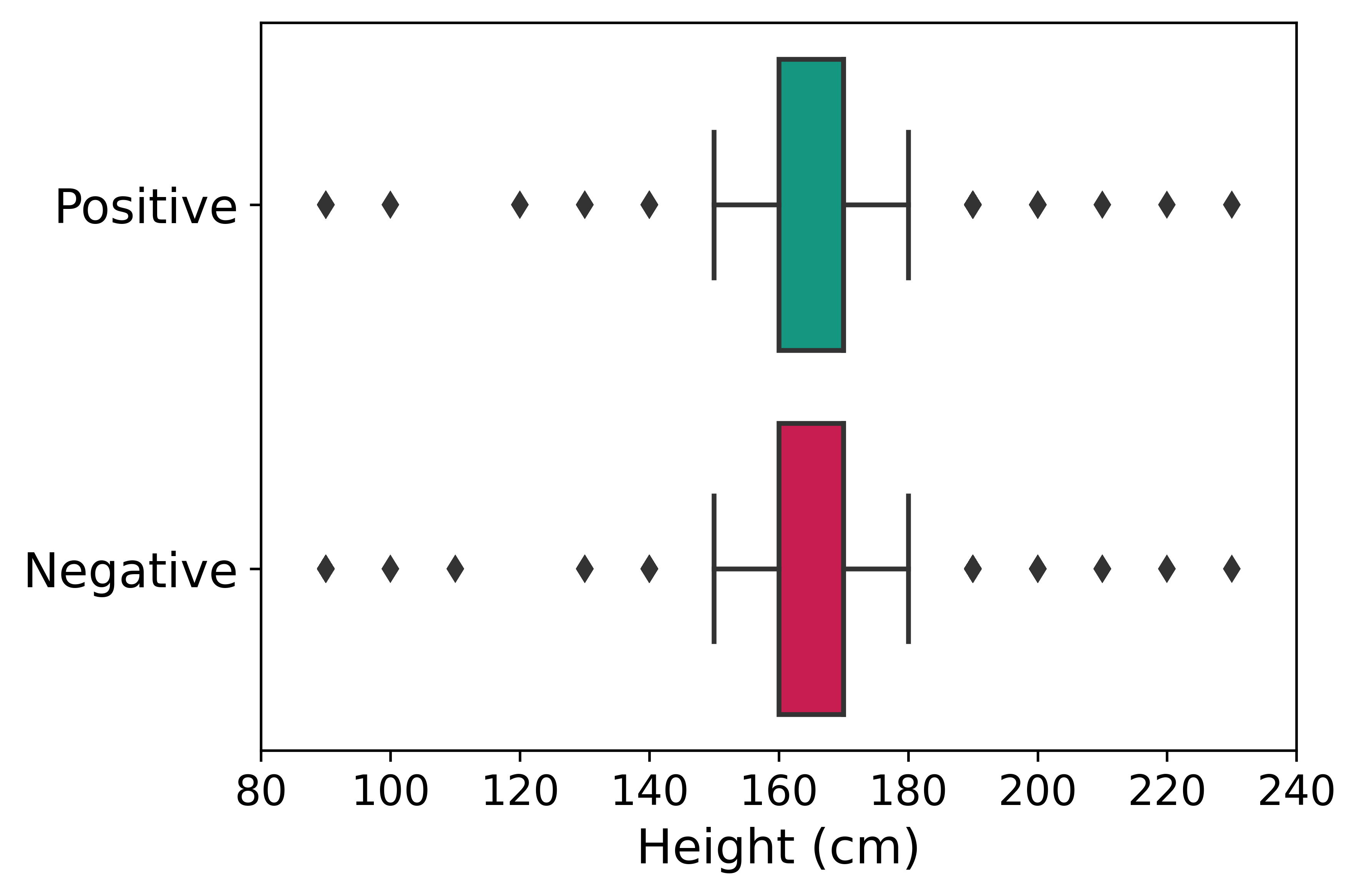}
    \includegraphics[scale=0.45]{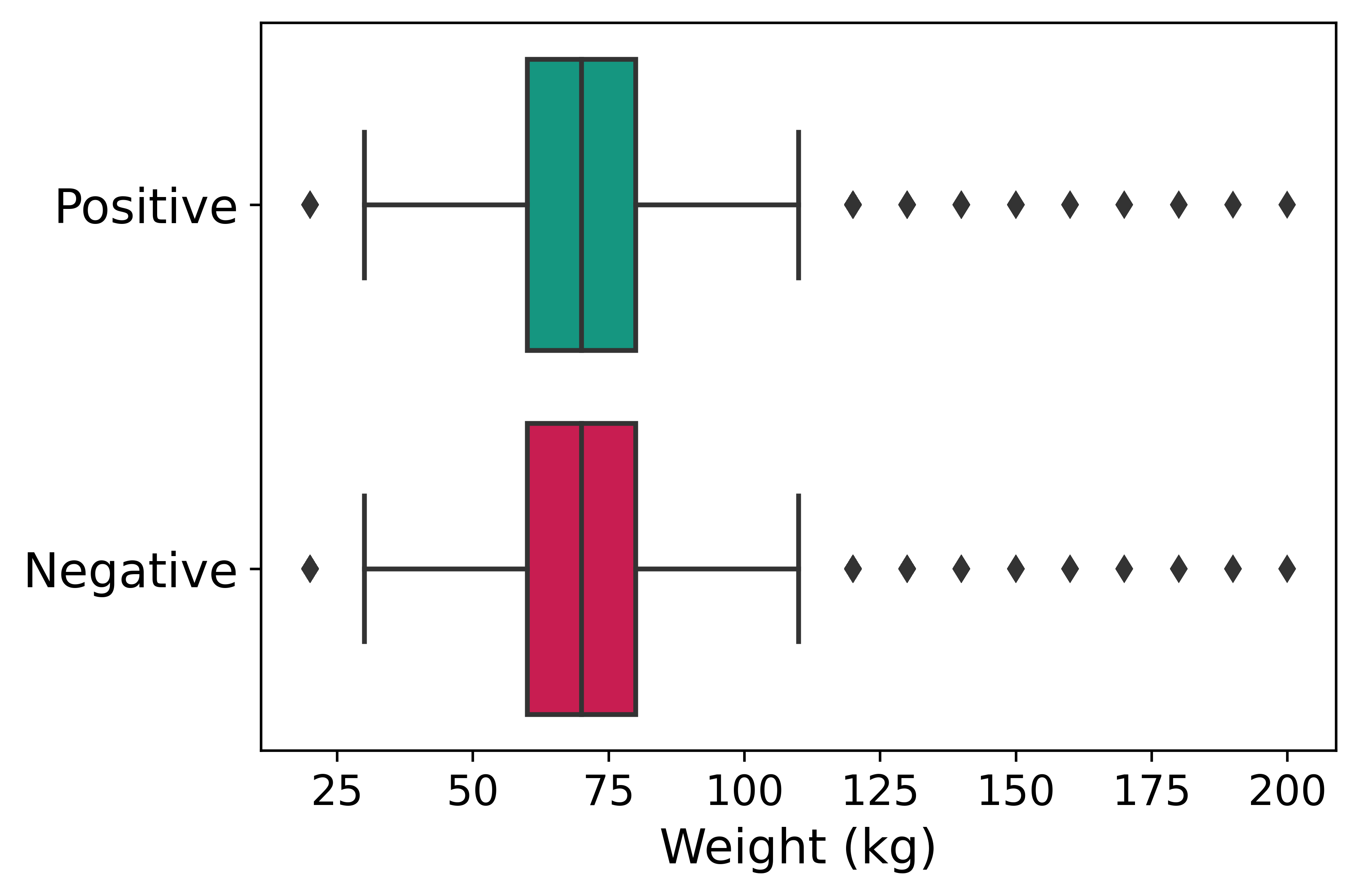}
    \caption{Left: Subjects' height broken down by COVID-19 status in the final dataset.  Right: Subjects' weight broken down by COVID-19 status in the final dataset.}
    \label{fig:covid_by_heigth}
\end{figure}

Since weight  and height were recorded only as classes, to analyse the height and weight data, the midpoints of the bins were taken and then converted to centimetres or kilograms for height and weight respectively. Height and weight do not appear to be confounded with COVID-19 status (see Figure \ref{fig:covid_by_heigth}), and the distributions of heights and weights appear reasonable, the only exception being a proportion of participants who selected the lowest height and weight option, probably due to these being the first options offered in the form \citep[see][for more details]{ciab_data}.  

\begin{figure}
    \centering
    \includegraphics[scale = 0.5]{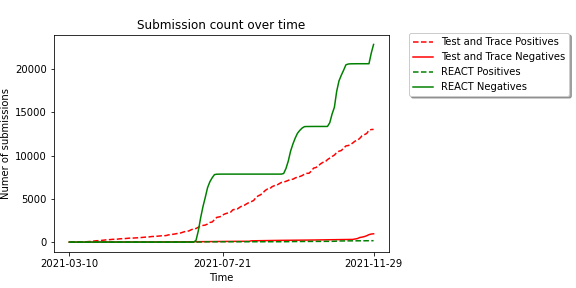}
    \caption{ \label{fig:submission} Submission count over time for both the Test and Trace and REACT recruitment channels.}
\end{figure}

\paragraph{Recruitment source.} Members of the public were invited to take part in the study after undergoing a test through two existing initiatives: NHS Test and Trace community testing programme (Pillar 2) and REACT-1 (community prevalence survey). The difference in these two testing initiatives meant that the NHS Test and Trace was a large recruiter of COVID-19 positive participants, since it was aimed at subjects suspected of having been infected and initially people were only contacted after a positive test result was confirmed \citep[see][]{ciab_data}. On the other hand, REACT-1 survey was aimed to randomly sample the population and hence, recruited a large number of negative cases with a few more positives coming through in periods of higher prevalence. Because of the recruitment structure, the recruitment channel is almost entirely confounded with COVID-19 infection status, as can be seen in Figure \ref{fig:submission} and Table \ref{tab:crosstab_tables}, which show the recruitment pattern over time and cross-tables between these variables respectively. Each rapid increase in REACT-1 submissions over time corresponds to the different REACT-1 recruitment rounds.

\begin{table}
\caption{\label{tab:crosstab_tables} Absolute frequency tables of COVID-19 status vs recruitment source, \\
COVID-19 status vs symptoms and symptoms vs recruitment source.}

\begin{tabular}{ccc}
    \begin{tabular}{c|c|c}
         & COV+ & COV-  \\ \hline
        T\&T & 13035 & 962 \\ \hline
        REACT & 164 & 22857
    \end{tabular} &
    \begin{tabular}{c|c|c}
         & COV+ & COV- \\ \hline
        Symp & 12753 & 5548 \\ \hline
        No Symp & 446 & 18271
    \end{tabular} &
    \begin{tabular}{c|c|c}
        & T\&T & REACT \\ \hline
        Symp & 13256 & 5045 \\ \hline
        No Symp & 741 & 17976
    \end{tabular}
\end{tabular}
\end{table}

\paragraph{Geographic dispersion.}

\begin{figure}[h!]
  \centering
  \includegraphics[width=0.9\linewidth]{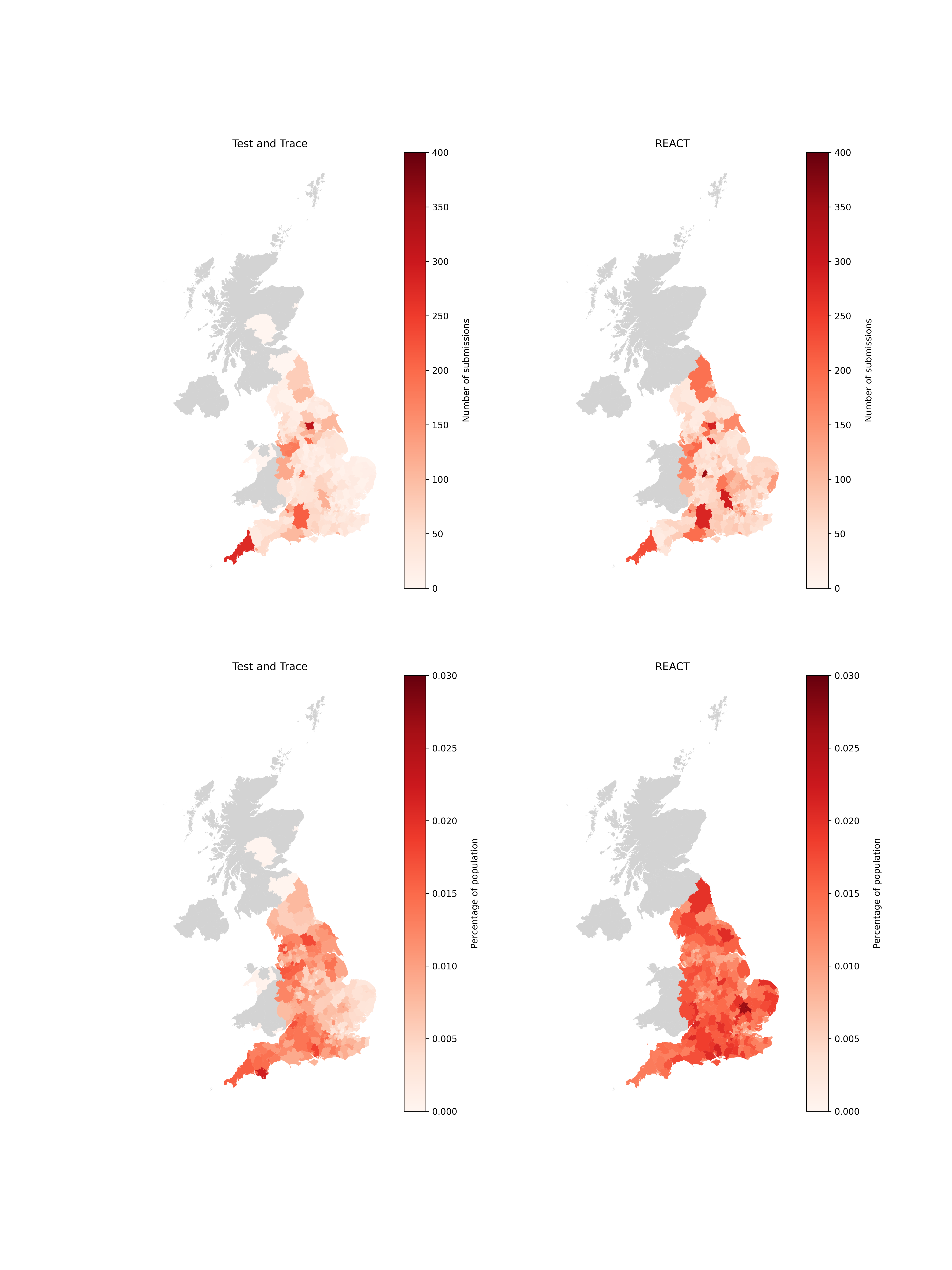}
  \caption{Geographic distribution of the submissions by recruitment source (Left:  NHS Test $\&$ TRACE, Right: REACT-1). Top: Total number of submissions. Bottom: Submissions as proportions of the total population.}
  \label{fig:geoplots}
\end{figure}
The geographic origin of the submissions for both recruitment sources can be seen in Figure \ref{fig:geoplots}. The geographic distribution of the submissions appears to be broadly similar for the two recruitment sources (and therefore for the two COVID-19 statuses, due to the confounding discussed above). On the other hand, not all regions contributed to the data collection in the same way and this is therefore identified as a potential source of confounding. It should also be noticed that the vast majority of the submissions come from England, with only a very small number of submissions coming from Scotland and Wales via the NHS Test and Trace route and none from Northern Ireland.  


\section{Choice of Challenging Test Sets}
\label{sec:choice}

Having performed the exploratory analysis of the metadata, our aim was to choose a split between train and test sets such that performance on the test set mimicked the performance on the general population as well as possible. For this purpose, variables which showed large imbalances between positive and negative cases were considered as potential confounders for audio-based classifiers. We considered, for example:

\begin{itemize}
    \item  Age
	\item Recruitment Source
	\item Symptoms (Cough (any), Fatigue, Headache, A change to sense of smell or taste, Runny or blocked nose, Fever (a high temperature), Loss of taste, Shortness of breath, Sore throat, A new continuous cough, Diarrhoea, Abdominal pain, Other symptom(s) new to you in the last 2 weeks, No symptoms, Prefer not to say.) All are binary variables -- either present (True) or not (False).
	\item Geographic location.
\end{itemize}

Secondly, the test set was chosen to systematically over-represent category combinations that are sparse in the overall data set. These are submissions with:
\begin{itemize}
    \item First language different from English. 
	\item Ethnicity different from White British.
	\item Older individual with positive tests results.
	\item Younger individual with negative test results.
\end{itemize}

We then proceed in defining a split between training and test sets of 25,897 and 11,121 submissions respectively, a $70\%$-$30\%$ train-test split, which was not uniformly randomly selected but based instead on the information coming from the meta-data analysis. The detailed procedure to sample the test set can be found in Algorithm \ref{alg:test}, and we refer to the test set produced by this procedure as the \emph{designed} test set. 

\begin{algorithm}
\caption{Test set construction. The sample sizes in parentheses reflect the test set that was generated for the methods' assessment.}\label{alg:test}
The following steps have been applied in this order to construct the test set from the submission meta-data:
\begin{enumerate}
\item 	Select all records from 5 randomly selected languages (excluding English) (n=370).
(to test out of sample performance for unseen first languages.)
\item	Select all records from 5 randomly selected ethnic or nationality groups (excluding British) (n=857).
(to test out of sample performance for unseen ethnic or nationality groups.)
\item	Select all negative cases from Leeds and Cornwall (n=547) (due to large numbers of positive submissions, to test out of sample performance for unseen locations) 
\item	Select all positive cases form Birmingham and Sheffield  (n=388) (due to large numbers of negative submissions, to test out of sample performance for unseen locations)
\item	Select all records from 4 other randomly selected local authorities (n=390)\\
(to test for geographic and dialectal confounding.)
\item	Select all asymptomatic cases (positive test result with ‘no symptoms’ selected) (n=439)\\
(to test if the audio-based method can be extended to asymptomatic positive cases, which are rare in the dataset.)
\item	Of those whose age is above the median by gender and tested positive, $50\%$ of records are selected (n=1299).
\item	Of those whose age is below the median by gender and tested negative, $50\%$ of records are selected (n=3032). \\
(to understand if Age could be acting as a confounding variable).
\item	Select all REACT positives (n=79)
\item	Select all Test and Trace negatives (n=962)\\
(to understand if recruitment channel could be acting as a confounding variable.)
\item	Sample without replacement to ensure there is an even distribution of viral load categories (n=598 high, 598 medium and 598 low)
(in case the analysis of accuracy of the models by viral load is required.)
\item	Fill test set to a 70-30 split by sampling without replacement from the remaining data randomly from records where viral load is not recorded (n=2932)
\end{enumerate}

Note that the groups listed above in (a)-(l) are not mutually exclusive. 

\end{algorithm}

However, in addition to the out-of-sample performance, there were also some concerns that the method could act as a symptoms (or more generally metadata variable) detector and any positive performance could be entirely due to the confounding, which was only partially accounted for in the designed test set and it would limit the usefulness of method as diagnostic tool.

Therefore, it is also appropriate to test the performance of the method on a subset of the designed test set, which has been constructed to control for all possible confounders in the collected data. We refer to this test set as the 1:1 \emph{matched} test set and it is constructed by exactly balancing the numbers of COVID-19 positive and COVID-19 negative individuals in each stratum, where to be in the same stratum individuals must be matched on all of the following variables: 
\begin{itemize}
    \item Recruitment channel (Test and Trace or REACT).
    \item Age (binned in 10 years intervals).
    \item Gender.
    \item Cough (TRUE or FALSE).
    \item Sore throat (TRUE or FALSE).
    \item Asthma (TRUE or FALSE).
    \item Shortness of breath (TRUE or FALSE).
    \item Runny/blocked nose (TRUE or FALSE).
    \item “At least one symptom” (TRUE or FALSE).
\end{itemize}

The matched test set has 984 COVID-19 positive and 984 COVID-19 negative participants. This reduces the size of the test set, resulting in a more noisy estimation of the prediction performance, but it also should remove the bias introduced by the known (recorded) confounders.

\section{Results of application of machine learning and statistical methods}
The detailed description of the AI methods used to predict COVID-19 status based on the audio submissions is beyond the scope of this paper and can be found in \cite{ciabdraft}, together with a full discussion of the results on the designed and matched test sets, based on multiple indicators. This paper reproduces a summary of the results of the analysis for one of the methods used, i.e., a support vector machine classifier based only on audio features extracted with open source speech and music interpretation by large-space extraction (openSMILE) \citep[see][for more details]{eyben2013recent}. We also compare it with a logistic regression classifier based only on the metadata. This is to have a benchmark of what would be the predictive accuracy of methods based on the meta-data instead of the acoustic features. By construction, this logistic regression will not be able to perform well on the matched test set. A random split between training and test sets is also considered for comparison, to see what we could have concluded in absence of the careful analysis we carried out. 

The area under the ROC curve index (AU-ROC) computed for the prediction on the designed and matched test set can be found in Table \ref{tab:AROC}. It can be seen that for both methods the performance is significantly 
worse for the matched test set, thus giving credence to the hypothesis that any signal in the data is due to the confounding variables. Furthermore, even on the designed test set, the performance of the method based on audio submission is worse than that of the logistic regression based on metadata, thus supporting the hypothesis that this method is not able to find additional information in the audio signals with respect to the metadata. Both methods perform well on the randomly selected test set, with performances in line with previously published studies discussed in Section \ref{sec:review}. 

\begin{table}
\caption{\label{tab:AROC} Area under the ROC Curve for the SVM classifier based on audio samples
and for the logistic regression based on the metadata, when applied to a randomised train-test split, the designed train-test split obtained with Algorithm \ref{alg:test},
and a matched test set obtained from the designed test-train split as described in Section \ref{sec:choice}, respectively.}
\begin{tabular}{ | c |c| c |c |}
\hline
& Randomised & Designed  &  Matched\\
\hline
 SVM - openSMILE & 0.80 & 0.73 & 0.60 \\ 
 \hline
 Logistic regression & 0.95 & 0.89 & 0.59 \\ 
 \hline
\end{tabular}
\end{table}

\section{Discussion}

Recently, there has been much interest in using machine learning methods to predict COVID-19 status based on audio signals, for example coughs. While early results seemed promising, they were affected by limitations in the data quality and their out-of-sample performance was an open issue. The `Speak up to help beat coronavirus' study, carried out by UKHSA, collected one of the largest and most comprehensive datasets available to date to explore this scientific question (the UK COVID-19 Vocal Audio Dataset). However, despite the quality and size of the collected data, biases are still present due to the inherent limitations in the data collection procedure, and particular care needs to be applied when designing a study to assess the out-of-sample performance of machine learning techniques. When confounding is accounted for with a matching strategy, the predictive power of the acoustic features becomes negligible in our model. This does not necessarily imply that it is not possible to use acoustic features to predict COVID-19 infection status, but we think we convincingly showed that without exceptional care in designing the assessment procedure the risk of reporting inflated performance results is very real, even with a large and well-curated dataset.

This study highlights important lessons on the use of machine learning techniques in observational studies in health sciences, that go beyond the immediate application to audio-based classifiers. Even with very large datasets and multiple data sources, biases and confounding variables can be present and dramatically overstate the prediction accuracy of machine learning methods. The estimated prediction accuracy can largely be corrected by using a matching strategy that approximates a case-control study, but this comes at the cost of a much reduced size for the test set (and potentially the training set, see \cite{ciabdraft} for more details on how this strategy can be applied to the training as well).

It is therefore important to design the data collection procedure to record participant information that can be associated with the outcome to be predicted. However, in the case of the ``Speak up and help beat Coronavirus" study, some of the confounding was clearly due to the data collection procedure. In particular, the much lower participation rate in the study for symptomatic individuals who tested negative via NHS Test and Trace, in comparison with the symptomatic individuals who tested positive. For future studies, this suggests the need to monitor at least the most obvious potential imbalances during the data collection and to intervene by making additional effort in recruiting underrepresented subgroups into the study. 

\section*{Ethics}
The study described in this paper has been approved by The National Statistician’s Data Ethics Advisory Committee (reference NSDEC(21)01) and the Cambridge South NHS Research Ethics Committee (reference 21/EE/0036) and Nottingham NHS Research Ethics Committee (reference 21/EM/0067). 

\section*{Data and code availability statement}
The data used in this paper are not publicly available. Access to the UK COVID-19 Vocal Audio Dataset may be requested from UKHSA (DataAccess@ukhsa.gov.uk), and will be granted subject to approval and a data sharing contract. To learn about how to apply for UKHSA data, visit: \url{https://www.gov.uk/government/publications/accessing-ukhsa-protected-data/accessing-ukhsa-protected-data}, see \cite{ciab_data} for more details. The code used to analyse the data and generate the train/test splits will be made available on the Alan Turing Institute GitHub repository.

\section*{Acknowledgments}
Authors in The Alan Turing Institute and Royal Statistical Society Health Data Lab gratefully acknowledge funding from Data, Analytics and Surveillance Group, a part of the UKHSA. This work was funded by The Department for Health and Social Care (Grant ref: 2020/045) with support from The Alan Turing Institute (EP/W037211/1) and in-kind support from The Royal Statistical Society.


\bibliographystyle{rss}
\bibliography{ciab}

\begin{thebibliography}{14}
\expandafter\ifx\csname natexlab\endcsname\relax\def\natexlab#1{#1}\fi
\expandafter\ifx\csname url\endcsname\relax
  \def\url#1{\texttt{#1}}\fi
\expandafter\ifx\csname urlprefix\endcsname\relax\def\urlprefix{URL: }\fi

\bibitem[{Babic et~al.(2021)Babic, Gerke, Evgeniou and Cohen}]{babic2021beware}
Babic, B., Gerke, S., Evgeniou, T. and Cohen, I.~G. (2021) Beware explanations
  from {AI} in health care.
\newblock \textit{Science}, \textbf{373}, 284--286.

\bibitem[{Brown et~al.(2020)Brown, Chauhan, Grammenos, Han, Hasthanasombat,
  Spathis, Xia, Cicuta and Mascolo}]{brown2020exploring}
Brown, C., Chauhan, J., Grammenos, A., Han, J., Hasthanasombat, A., Spathis,
  D., Xia, T., Cicuta, P. and Mascolo, C. (2020) Exploring automatic diagnosis
  of {COVID}-19 from crowdsourced respiratory sound data.
\newblock \textit{arXiv preprint arXiv:2006.05919}.

\bibitem[{Budd et~al.(2022)Budd, Baker, Karoune, Coppock, Patel, Tendero
  Ca\~{n}adas, Titcomb, Payne, Hurley, Egglestone, Butler, Mellor, Nicholson,
  Kiskin, Koutra, Jersakova, McKendry, Diggle, Richardson, Schuller, Gilmour,
  Pigoli, Roberts, Packham, Thornley and Holmes}]{ciab_data}
Budd, J., Baker, K., Karoune, E., Coppock, H., Patel, S., Tendero Ca\~{n}adas,
  A., Titcomb, A., Payne, R., Hurley, D., Egglestone, S., Butler, L., Mellor,
  J., Nicholson, G., Kiskin, I., Koutra, V., Jersakova, R., McKendry, R.,
  Diggle, P., Richardson, S., Schuller, B., Gilmour, S., Pigoli, D., Roberts,
  S., Packham, J., Thornley, T. and Holmes, C. (2022) A large-scale and
  pcr-referenced vocal audio dataset for {COVID}-19.
\newblock \textit{arXiv preprint}.

\bibitem[{Coppock et~al.(2021)Coppock, Jones, Kiskin and
  Schuller}]{coppock2021covid}
Coppock, H., Jones, L., Kiskin, I. and Schuller, B. (2021) Covid-19 detection
  from audio: seven grains of salt.
\newblock \textit{The Lancet Digital Health}, \textbf{3}, e537--e538.

\bibitem[{Coppock et~al.(2022)Coppock, Nicholson, Kiskin, Koutra, Baker, Budd,
  Payne, Karoune, Hurley, Titcomb, Egglestone, Tendero Ca\~{n}adas, Butler,
  Jersakova, Patel, Thornley, Mellor, Diggle, Richardson, Packham, Schuller,
  Gilmour, Pigoli, Roberts and Holmes}]{ciabdraft}
Coppock, H., Nicholson, G., Kiskin, I., Koutra, V., Baker, K., Budd, J., Payne,
  R., Karoune, E., Hurley, D., Titcomb, A., Egglestone, S., Tendero
  Ca\~{n}adas, A., Butler, L., Jersakova, R., Patel, S., Thornley, T., Mellor,
  J., Diggle, P., Richardson, S., Packham, J., Schuller, B., Gilmour, S.,
  Pigoli, D., Roberts, S. and Holmes, C. (2022) Audio-based {AI} classifiers
  show no evidence of improved {COVID}-19 screening over simple symptoms
  checkers.
\newblock \textit{arXiv preprint}.

\bibitem[{Eyben et~al.(2013)Eyben, Weninger, Gross and
  Schuller}]{eyben2013recent}
Eyben, F., Weninger, F., Gross, F. and Schuller, B. (2013) Recent developments
  in opensmile, the munich open-source multimedia feature extractor.
\newblock In \textit{Proceedings of the 21st ACM international conference on
  Multimedia}, 835--838.

\bibitem[{Han et~al.(2021)Han, Brown, Chauhan, Grammenos, Hasthanasombat,
  Spathis, Xia, Cicuta and Mascolo}]{han2021exploring}
Han, J., Brown, C., Chauhan, J., Grammenos, A., Hasthanasombat, A., Spathis,
  D., Xia, T., Cicuta, P. and Mascolo, C. (2021) Exploring automatic {COVID}-19
  diagnosis via voice and symptoms from crowdsourced data.
\newblock In \textit{ICASSP 2021-2021 IEEE International Conference on
  Acoustics, Speech and Signal Processing (ICASSP)}, 8328--8332. IEEE.

\bibitem[{Han et~al.(2022)Han, Xia, Spathis, Bondareva, Brown, Chauhan, Dang,
  Grammenos, Hasthanasombat, Floto et~al.}]{han2022sounds}
Han, J., Xia, T., Spathis, D., Bondareva, E., Brown, C., Chauhan, J., Dang, T.,
  Grammenos, A., Hasthanasombat, A., Floto, A. et~al. (2022) Sounds of
  {COVID}-19: exploring realistic performance of audio-based digital testing.
\newblock \textit{NPJ digital medicine}, \textbf{5}, 1--9.

\bibitem[{Laguarta et~al.(2020)Laguarta, Hueto and
  Subirana}]{laguarta2020covid}
Laguarta, J., Hueto, F. and Subirana, B. (2020) {COVID}-19 artificial
  intelligence diagnosis using only cough recordings.
\newblock \textit{IEEE Open Journal of Engineering in Medicine and Biology},
  \textbf{1}, 275--281.

\bibitem[{Riley et~al.(2020)Riley, Atchison, Ashby, Donnelly, Barclay, Cooke,
  Ward, Darzi, Elliott, Group et~al.}]{riley2020real}
Riley, S., Atchison, C., Ashby, D., Donnelly, C.~A., Barclay, W., Cooke, G.~S.,
  Ward, H., Darzi, A., Elliott, P., Group, R.~S. et~al. (2020) Real-time
  assessment of community transmission ({REACT}) of {SARS-CoV}-2 virus: study
  protocol.
\newblock \textit{Wellcome Open Research}, \textbf{5}.

\bibitem[{Rudin(2019)}]{rudin2019stop}
Rudin, C. (2019) Stop explaining black box machine learning models for high
  stakes decisions and use interpretable models instead.
\newblock \textit{Nature Machine Intelligence}, \textbf{1}, 206--215.

\bibitem[{{UK Health Security Agency}(2021)}]{linktodhscpage}
{UK Health Security Agency} (2021) Speak up and help beat coronavirus.
\newblock
  \url{https://www.gov.uk/government/news/speak-up-and-help-beat-coronavirus-covid-19}.

\bibitem[{Watson(2022)}]{watson2022conceptual}
Watson, D.~S. (2022) Conceptual challenges for interpretable machine learning.
\newblock \textit{Synthese}, \textbf{200}, 1--33.

\bibitem[{{World Health Organization}(2020)}]{who2020}
{World Health Organization} (2020) {WHO Director-General's opening remarks at
  the media briefing on COVID-19 - 16 March 2020}.
\newblock
  \url{https://www.who.int/director-general/speeches/detail/who-director-general-s-opening-remarks-at-the-media-briefing-on-covid-19---16-march-2020}.
\newblock Accessed: 15 September 2022.

\end{thebibliography}
\end{document}